\documentclass[twocolumn,printnumbers,amsmath,amssymb,showpacs,prl,longbibliography]{revtex4-2}
\usepackage{graphicx}
\usepackage{colordvi}
\usepackage{color}
\usepackage[british]{babel}

\makeatletter
\def\fnum@figure#1{FIG.~\thefigure$:$~}
\makeatother

\begin{document}

\title{{Universal Jamming Criticality and Self-Organizing Principles from Disorder to \\
the Limit of Perfect Crystalline Order}}

\author{Jianhua Zhang$^{1}$}
\author{Jiaqi Si$^{1,2}$}
\author{Ning Xu$^{1,2}$}
\author{Hua Tong$^{1,}$}
\email[Email: ]{huatong@ustc.edu.cn}
\affiliation{$^1$Department of Physics and Anhui Center for Fundamental Sciences in Theoretical Physics, University of Science and Technology of China, Hefei 230026, China \\
$^2$Hefei National Laboratory for Physical Sciences at the Microscale and CAS Key Laboratory of Microscale Magnetic Resonance, University of Science and Technology of China, Hefei 230026, China}
\date{\today}

\begin{abstract}
While crystals are defined by periodic order, the nature of amorphous solids remains elusive due to their disordered, diverse, and nonequilibrium structures. {Here, we focus on jammed elastic packings and systematically tune structure from crystalline to fully disordered to unveil the universal underlying characteristics.} We demonstrate that their mechanical properties are universally governed by jamming criticality, featuring characteristic scaling behaviors near the jamming transition, excepting the singular close-packed point. This is facilitated by random nonaffine elasticity arising from contact-level disorder. Consequently, the jamming density can approach close packing, suggesting a fundamental decoupling between jamming criticality and the glass transition physics. Moreover, we uncover a universal coordination-number distribution and contact hyperuniformity in marginally jammed states, independent of particle-level structure. These findings suggest a general organizing mechanism for emergent rigidity in disordered solids, underscore the broad relevance of jamming physics, and complement principles of mechanical self-organization.
\end{abstract}
			
\maketitle
{\it Introduction.} -- All crystals are alike, but each amorphous solid is disordered in its own way~\cite{ashcroft1976solid,binder2011glassy}. While translational periodicity enables robust theoretical descriptions of crystals, amorphous solids -- including diverse systems from molecular glasses to granular matter, colloids, and foams -- lack such a unified structural basis~\cite{phillips1981amorphous,binder2011glassy}. Jammed packings of elastic particles have been a fertile playground for elucidating common features of amorphous solids, providing insights into universal glassy phenomena from elastoplasticity to low-temperature anomalies \cite{liu2010jamming,van2009jamming,torquato2010jammed,bi2015statistical}. Although initially focused on random close packings, jamming physics has recently proven effective in describing the mechanics of highly ordered solids \cite{goodrich2014solids,tong2015crystals,charbonneau2019glassy,tsekenis2021jamming,ikeda2020jamming}; however, its exact range of applicability remains elusive. Here, we address this problem by characterizing jamming criticality across the spectrum of structural order, seeking a unified understanding of the organizing principles in marginally jammed solids.

In its purest form, jamming transition occurs at zero temperature when particles get stuck upon quasistatic densification and gain rigidity \cite{liu2010jamming,torquato2010jammed,van2009jamming}. While the physics of jamming carries on at finite temperatures or during dynamic processes driven by shear or compression \cite{zhang2009thermal,degiuli2015theory,ikeda2013dynamic,heussinger2009jamming,olsson2019dimensionality,peshkov2021critical,peshkov2022universality}, we restrict our focus to the classical zero-temperature jamming in this work. According to the Maxwell criterion, this happens when the average coordination number reaches isostaticity, $Z_c=2d$ ($d$ is the spatial dimension).
This resembles rigidity percolation, but the jamming transition is unique~\cite{ellenbroek2009non,ellenbroek2015rigidity}. In the case of harmonic interaction, the coordination number $Z$ and bulk modulus $B$ jump discontinuously, whereas the shear modulus $G$ grows continuously \cite{o2003jamming}.
By contrast, all these quantities grow continuously for standard rigidity percolation \cite{feng1985effective,jacobs1995generic}. 
This distinction implies nontrivial structural self-organization in jammed solids beyond mean-field isostaticity, driven by strong steric constraints \cite{ellenbroek2009non,ellenbroek2015rigidity,hagh2019broader,bi2015statistical}. While lattice models incorporating compression-resistant subsets have been proposed to explain this mixed character \cite{liarte2019jamming}, their correspondence to apparently disordered solids is unclear. Revealing the hidden structural characteristics is thus crucial for understanding the distinct mechanism of emergent rigidity in jammed solids.

Mean-field (MF) approaches, like the replica theory of hard-sphere glasses in infinite dimensions, have provided valuable insights into the nature of marginally jammed solids \cite{parisi2010mean,kurchan2012exact,charbonneau2017glass}. It predicts a Gardner transition into a marginal state upon compression, which explains the marginal stability of jammed solids \cite{charbonneau2014fractal,charbonneau2017glass,kurchan2012exact}. It sets a maximum jamming packing fraction $\phi_{\rm GCP}$, corresponding to jammed states compressed from ideal glasses. When generalized to physical dimensions, it predicts $\phi_{\rm GCP}=0.8745$ (2D) and $0.6836$ (3D) \cite{parisi2010mean}. However, strong steric constraints in low dimensions may introduce nonperturbative many-body correlations \cite{tanaka2012bond,tanaka2019revealing}, potentially altering these MF predictions. Meanwhile, variational arguments and effective medium theory applied to soft-particle packings yield consistent results with replica theory \cite{wyart2005effects,wyart2005rigidity,wyart2010scaling,mao2010soft,degiuli2014effects}. However, traditional effective medium theory fails to capture the critical scaling of $B$ and $G$ near jamming because it averages out crucial spatial fluctuations \cite{wyart2010scaling,feng1985effective,makse1999effective}.
Therefore, identifying underlying constraints in mechanical network is essential for accurate theoretical descriptions in physical dimensions.

In this Letter, we investigate how disorder dictates the mechanical nature of marginally jammed  packings. By continuously modulating structural order, we realize packings from near-crystalline to fully disordered and reach two major findings. First, jamming criticality universally dominates below a critical pressure $p_j$ for any finite polydispersity $\eta$, with $p_j \sim \eta^{2.5}$ (2D) and $p_j \sim \eta^2$ (3D) in the weak-disorder regime.
This implies: (i) No finite-disorder threshold separates crystal and jamming behaviors; instead, the domain of jamming physics scales with disorder; (ii) The jamming density can approach close packing, significantly extending the J-line beyond MF predictions related to glass transition; (iii) Random nonaffine elasticity underlies the jamming behavior, controlled by contact-level disorder. Second, we find a universal coordination-number distribution and contact hyperuniformity in marginally jammed packings, regardless of particle-level structure. These results strongly support a unified definition and understanding of amorphous solids based on their universal mechanical self-organization.

{\it Methods.} -- Jammed packings are generated by quasi-statically increasing polydispersity $\eta$ from perfect crystals (hexagonal in 2D and face-centered cubic in 3D) \cite{mizuno2013elastic,mizuno2014acoustic,tong2015crystals}.
Here $\eta$ controls the width of the particle-size distribution and in turn the degree of structural disorder (see End Matter for details).
This modulates structural order continuously from near-crystalline to fully disordered while maintaining overall homogeneity, ensuring our systems are {\it pure phases} according to Gibbs definition rather than mixtures of ordered and disordered components \cite{gibbs1878equilibrium}.
Randomly jammed binary packings are studied as a reference \cite{liu2010jamming,o2003jamming}. Particles interact via harmonic or Hertzian repulsion. The distance to jamming transition is then controlled by pressure $p$ at fixed $\eta$.
All quantities are expressed in reduced units (see End Matter for detailed definitions).
To overcome the tremendous slowing down in the weak-disorder limit, a hybrid FIRE algorithm with self-activation and swaps is used to generate truly stable configurations \cite{bitzek2006structural}. More details are in End Matter. Unless noted, we present results for 2D harmonic systems with uniform particle-size distributions. Results from 2D binary and polydisperse systems with Gaussian particle-size distributions, and 2D Hertzian systems are also included in Fig.~\ref{fig3}(b), and those from 3D harmonic systems are given in Supplemental Material to validate the generality of our findings \cite{supplementary}.

\begin{figure}[ht]
\includegraphics[width=0.49\textwidth]{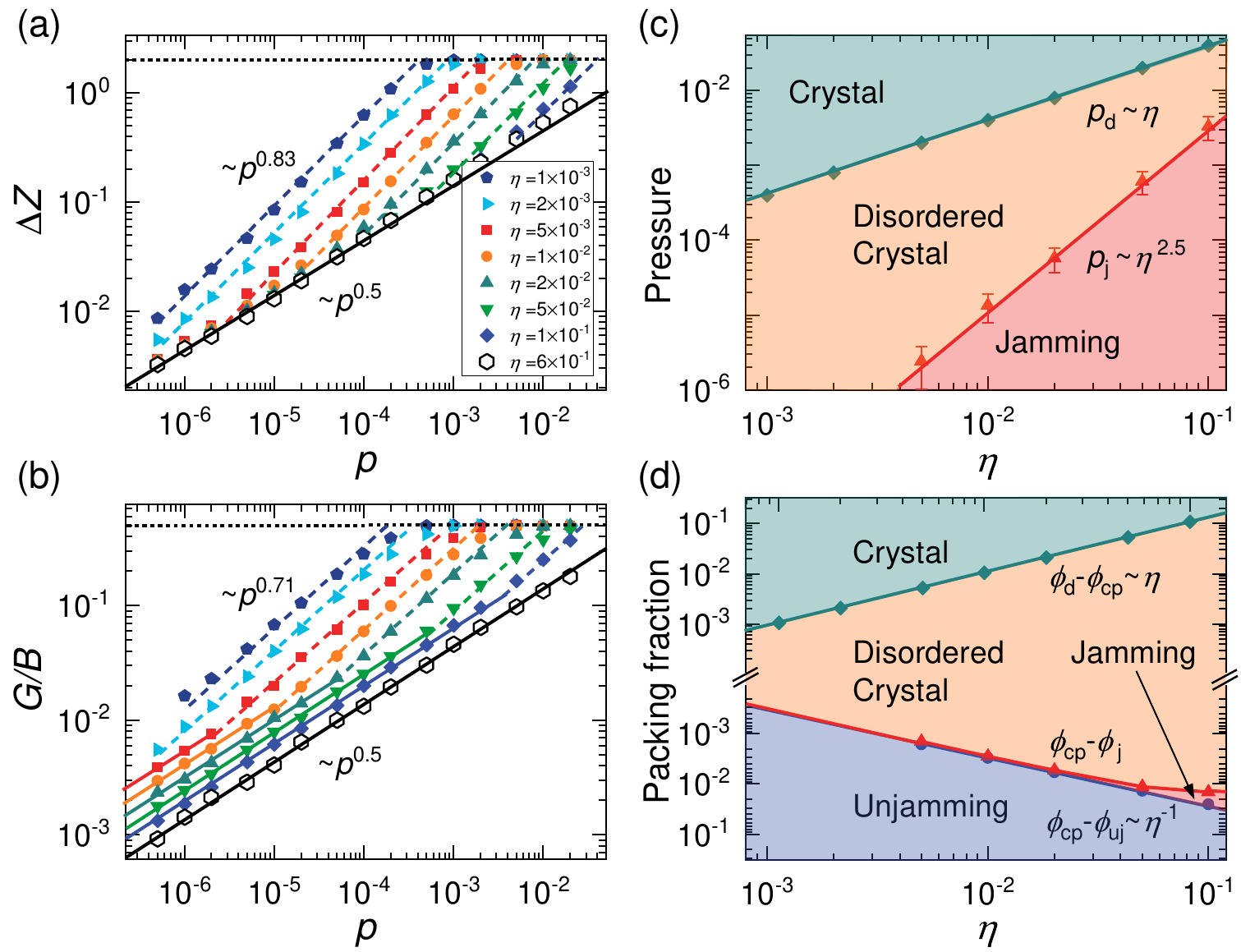}
\vspace{-0.15 in}
\caption{Jamming criticality in solids from order to disorder. Pressure dependence of (a) the excess coordination number $\Delta Z$ and (b) the modulus ratio $G/B$ for varied polydispersity $\eta$.  Horizontal dotted lines indicate crystal behavior. Dashed and solid lines denote intermediate scalings ($\Delta Z \sim p^{0.83}$, $G/B \sim p^{0.71}$) and jamming scalings ($\Delta Z \sim p^{0.5}$, $G/B \sim p^{0.5}$), respectively. Crossover pressures $p_d$ and $p_j$ are determined by intersections of different scaling relations. (c) Phase diagram in ($\eta$, $p$) space. Three regions with distinct mechanical behaviors are identified, separated by $p_d\sim \eta$ (green line) and $p_j \sim \eta^{2.5}$ (red line). Error bars indicate uncertainty of crossover pressures determined by $\Delta Z$ and $G/B$. (d) Phase diagram in ($\eta, |\phi - \phi_{\rm{cp}}|$) space, showing the unjammed region below $\phi_{\rm{uj}}$. Crossover packing fractions $\phi_d$ and $\phi_j$ correspond to $p_d$ and $p_j$, respectively. }
\label{fig1}
\end{figure}

{\it Phase diagram.} -- 
Figures~\ref{fig1}(a) and \ref{fig1}(b) show the pressure dependence of excess coordination number $\Delta Z=Z-Z_c$ and $G/B$ across a wide range of polydispersity $\eta$, including near-crystalline ($\eta \le 0.1$) to fully disordered structures ($\eta=0.6$) \cite{tong2015crystals}. Three characteristic regimes separated by two crossovers emerge for near-crystalline packings:
(i) Crystal regime at high pressures, with nearly constant $\Delta Z$ and $G/B$; (ii) Disordered crystal regime below $p_d$, with intermediate scalings $\Delta Z \sim p^\beta$ and $G/B \sim p^\gamma$, where $\beta \approx 0.83$ and $\gamma \approx 0.7$ in both 2D and 3D; (iii) Jamming critical regime below $p_j$, with universal jamming scalings $\Delta Z\sim p^{0.5}$ and $G/B\sim p^{0.5}$.
The first crossover at $p_d \sim \eta$ corresponds to a hidden transition reported previously from crystals to disordered crystals, which are solids with  extremely high crystalline order in structure but mechanical and vibrational properties resembling amorphous solids \cite{tong2015crystals}. Crucially, we identify $p_j$ as the second hidden crossover where solids become marginally stable and governed by jamming criticality. We find $p_j\sim \eta^\delta$, with $\delta=2.5$ (2D) and $\delta=2$ (3D). This demonstrates the applicability of jamming physics even at infinitely small $\eta$, refuting a finite-disorder threshold separating crystal and jamming behaviors. {The close packed point is therefore a singular point where the mechanical properties are controlled by crystal physics upon jamming. } Our result is consistent with a recent MF model of disordered crystals \cite{ikeda2020jamming}, though the 2D exponent $\delta=2.5$ deviates from MF prediction $\delta=2$.
While $d=2$ has been suggested as the upper critical dimension of the jamming transition \cite{goodrich2012finite,wyart2005rigidity}, our result echoes previous observations of dimensional differences in jamming criticality. In particular, different scaling behaviors of correlation lengths \cite{hexner2018two} and diverging viscosity driven by shear or compression \cite{olsson2019dimensionality,peshkov2021critical} have been reported in 2D and 3D.

\begin{figure}[t]
\includegraphics[width=0.48\textwidth]{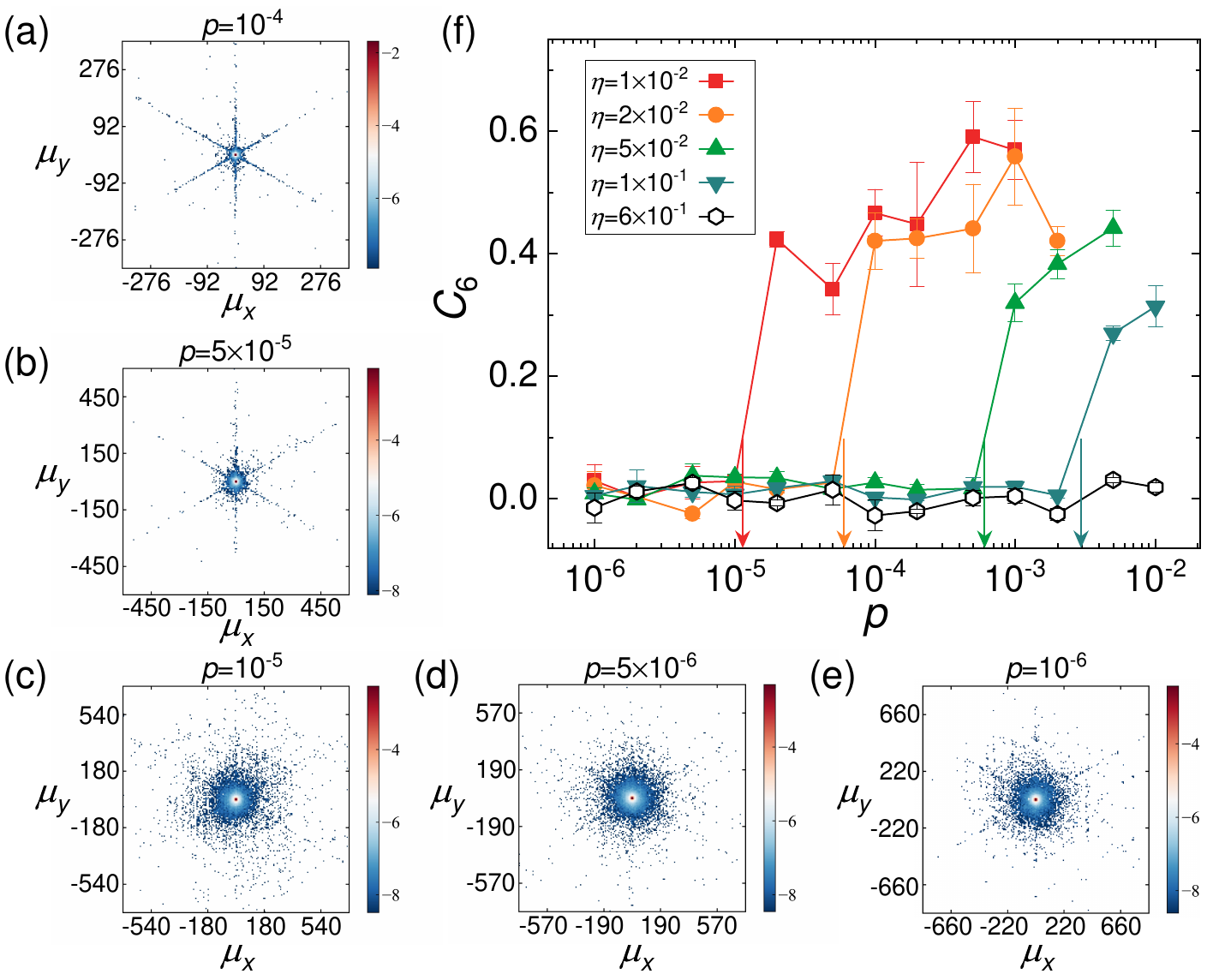}
\vspace{-0.15 in}
\caption{Nonaffinity in solids from order to disorder. (a)-(e) Probability distribution of normalized nonaffine deformation $p(\mu_x, \mu_y)$ for $\eta = 10^{-2}$ with decreasing pressures. (f) Six-fold anisotropy $C_6$ of $p(\mu_x, \mu_y)$ versus pressure $p$ for different $\eta$. Arrows indicate $p_j$, below which the system is controlled by jamming criticality. }
\label{fig2} 		
\end{figure}

The resulting phase diagram for weakly disordered solids is given in Fig.~\ref{fig1}(c), with three characteristic regimes highlighted in the space of polydispersity, i.e., the degree of disorder, and pressure. The conjugate phase diagram in the space of polydispersity and packing fraction is shown in Fig.~\ref{fig1}(d), with additional information of (un)jamming packing fraction $\phi_{\rm uj}$. Both $\phi_{\rm uj}$ and $\phi_j$ approach close packing $\phi_{\rm cp}$ as $\eta \to 0$, pushing the jamming packing fraction and the relevance of jamming physics to $\phi_{\rm cp}$. This significantly extends the J-line beyond previous numerical results \cite{chaudhuri2010jamming,ozawa2017exploring} and replica theoretical predictions \cite{parisi2010mean} based on amorphous packings.

Therefore, particle-level glassy (noncrystalline) structure is not a prerequisite for jamming criticality, suggesting that jamming physics can be essentially decoupled from the physics of glass transition. Instead, contact-level disorder, here controlled by polydispersity, is expected to determine the mechanics of marginally jammed solids via nontrivial self-organization \cite{xing2024origin,corwin2005structural}.
Our model with near-crystalline geometry thus provides a unique platform to reconsider the defining character of amorphous solids and unveil universal organizing principles across the entire order-disorder spectrum \cite{mari2009jamming,acharya2020athermal,lerner2022disordered,maharana2024universal}.

\begin{figure*}[ht]
\includegraphics[width=0.9\textwidth]{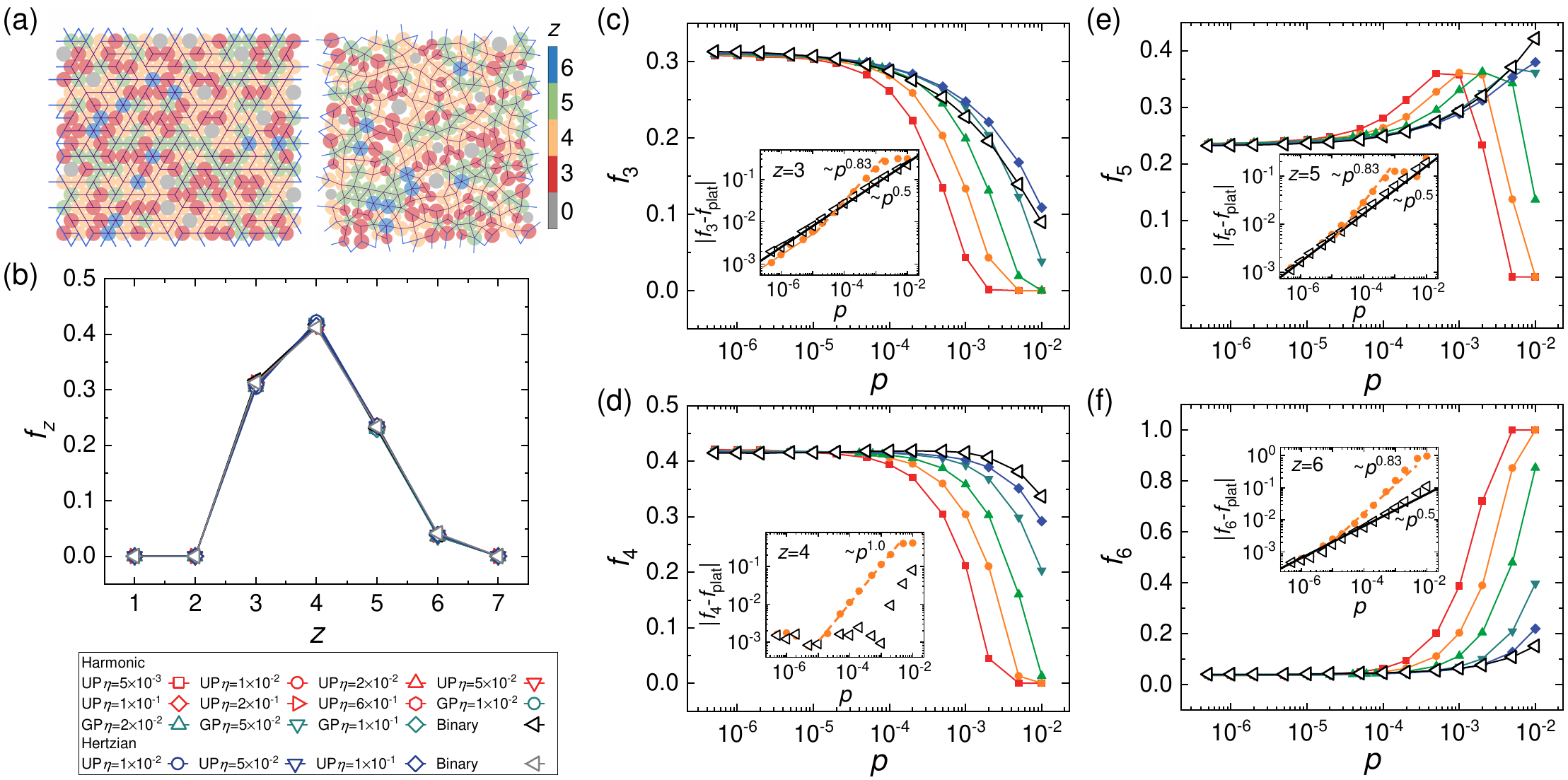}
\vspace{-0.15 in}
\caption{Universal coordination-number distribution in marginally jammed solids. (a) Representative configurations at $p=10^{-6}$ for polydisperse crystalline ($\eta = 10^{-2}$, left) and randomly quenched binary (right) systems. Colors denote the coordination number $z$ and force network is plotted on top. (b) Universal coordination-number distribution $f_{z}$ at the jamming transition in polydisperse systems derived from perfect crystal with uniform (UP) and Gaussian (GP) particle-size distributions, and randomly quenched binary system (Binary), with harmonic or Hertzian interactions. (c)-(f) Pressure dependence of $f_{z}$ for UP (from red to blue, $\eta = 5 \times 10^{-3}$, $10^{-2}$, $2 \times 10^{-2}$, $5 \times 10^{-2}$, and $10^{-1}$) and binary (black) systems. Insets: Pressure dependence of $| f_{z} - f_{\rm{plat}}|$ for UP ($\eta = 10^{-2}$, orange) and binary (black)  systems, with $f_{\rm{plat}}$ being the corresponding plateau value of $f_z$ approaching jamming transition. Dashed and solid lines indicate power-law scalings for UP at intermediate pressures and jamming behavior, respectively.}\label{fig3}
\end{figure*}

{\it Nonaffine elastic response.} -- The consistent jamming scalings in solids across near-crystalline to fully disordered structures imply a general underlying mechanism of elastic responses. This we analyze via nonaffine deformation crucial for critical jamming scalings \cite{ellenbroek2006critical,ellenbroek2009non}.
We define vectorial nonaffinity as the nonaffine displacement normalized by the affine part to resolve the influence of underlying geometric structures $\vec{\mu}_{i} = \delta \vec{r}_{i}^{\rm{NA}} / \sqrt{\sum_{j} \left(\delta \vec{r}_{j}^{\rm{A}} \right)^{2}/N}$, where $\delta \vec{r}_{i}^{\rm{NA}}$ and $\delta \vec{r}_{i}^{\rm{A}}$ are the nonaffine and affine displacements of particle $i$ upon compression, respectively \cite{tong2014order,supplementary}. Figures \ref{fig2}(a)-\ref{fig2}(e) show the probability distribution of $\vec{\mu}_{i}$ for near-crystalline packings ($\eta=10^{-2}$) as pressure decreases. At high pressures,
the distribution exhibits clear six-fold symmetry due to strong constraints by the underlying lattice. However,  around and below $p_j\approx 10^{-5}$, the distribution becomes essentially isotropic, indicating that the system responds flexibly, free from the particle-level structure.

Quantitatively, we define $C_6$ to measure the six-fold anisotropy of the distribution (see Supplemental Material \cite{supplementary} for definition and discussions). $C_6=0$ describes isotropic distributions, whereas $C_6=1$ indicates a perfect six-fold symmetry and therefore a strong influence from the triangular  lattice. As shown in Fig. \ref{fig2}(f), $C_6$ drops sharply near $p_j$ (indicated by arrows) for near-crystalline packings ($\eta\le 0.1$), while maintaining almost zero across the whole pressure range for fully disordered packings ($\eta=0.6$).
The crossover $p_j$ thus corresponds to a mechanism change of elastic responses: once the mechanical environment (overlaps and gaps between neighbours), as determined by the interplay of polydispersity and pressure, is sufficiently disordered at the contact level, the elastic response becomes dominated by random nonaffinity, regardless of particle-scale geometric order.
We further confirm from the static structure factor that the particle-level structure is irrelevant for the observed crossover into the jamming critical regime (see Supplemental Material \cite{supplementary}).

{\it Universal coordination-number distribution.} --  The universal jamming behavior observed in jammed packings with near-crystalline to fully disordered structures [Fig.~\ref{fig3}(a)] poses a stringent question: what hidden feature unifies these apparently distinct geometries? Characterizing the coordination network of the mechanical structure, we discover a universal probability distribution of the coordination number $f_z$ for marginally jammed solids as $p \to 0$, as shown in Fig.~\ref{fig3}(b). For all 2D harmonic systems, we find $f_{3} \approx 0.308$, $f_{4} \approx 0.416$, $f_{5} \approx 0.237$, and $f_{6} \approx 0.039$. We confirm the same result for 2D Hertzian systems, while 3D harmonic systems exhibit a weaker form of a universal distribution \cite{supplementary}, possibly due to weaker steric constraints in higher dimensions.
Our result thus unveils an additional constraint on the mechanical network for achieving rigidity \cite{ellenbroek2015rigidity}, complementing the global Maxwell criterion ($Z_c=2d$). This should arise from strong steric constraints in low-dimensional jammed configurations, which are crucially absent in MF models, explaining the observed deviations of scaling exponents in 2D from MF predictions \cite{ikeda2020jamming}. The absence of such constraints in typical lattice models may also underlie the distinct nature of rigidity percolation compared to jamming transition \cite{ellenbroek2009non,ellenbroek2015rigidity,liarte2019jamming}.


\begin{figure}[ht]
\includegraphics[width=0.49\textwidth]{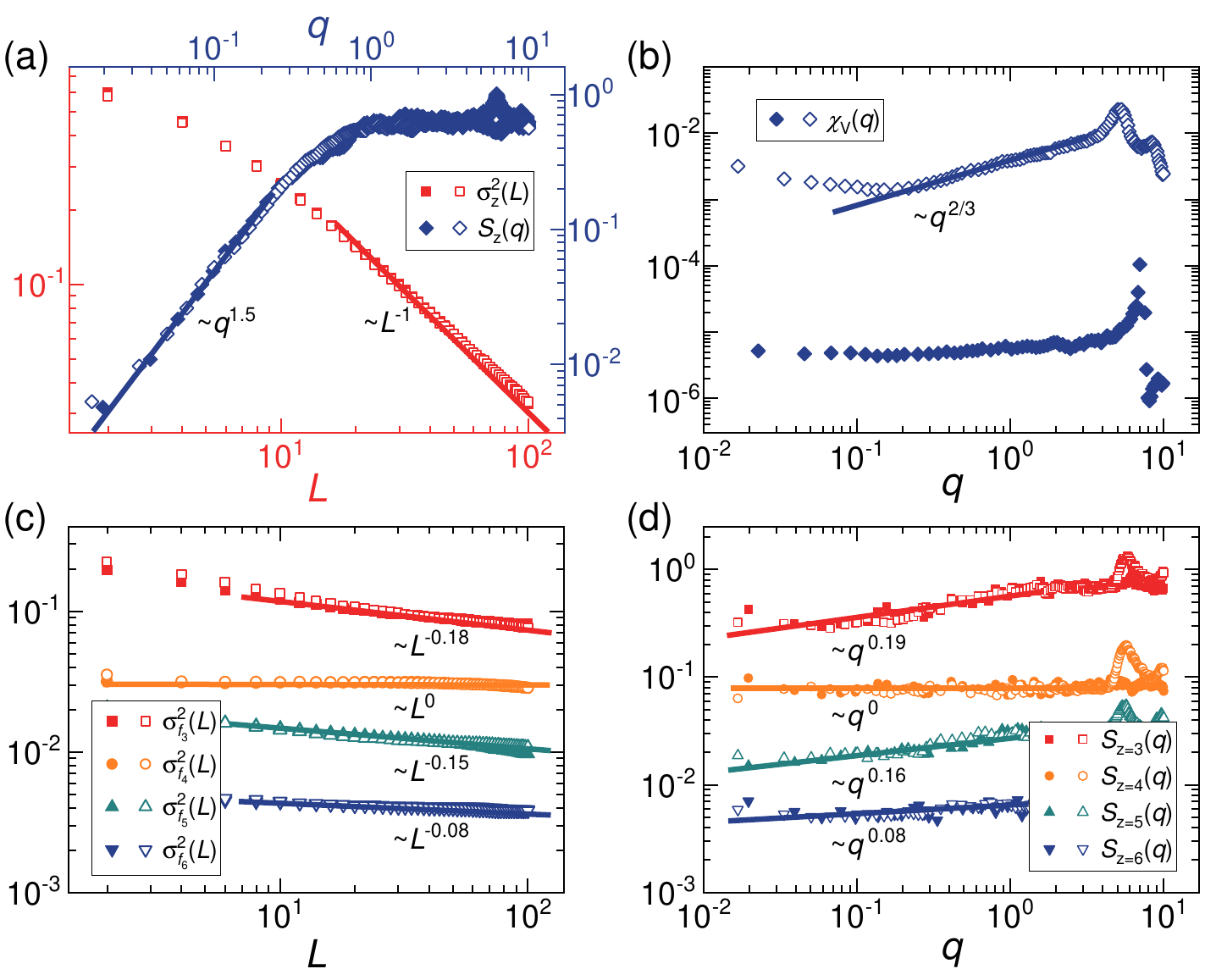}
\vspace{-0.15 in}
\caption{Suppressed spatial fluctuation of contacts and density. Comparison between near-crystalline ($\eta=10^{-2}$,  full symbols) and quenched binary systems (open symbols) at $p=10^{-6}$. (a) Real-space fluctuation $\sigma^2_{Z}(L)$ and structure factor $S_Z(q)$ of the coordination number, and (b) the corresponding spectral density $\chi_{V}({q})$ for the whole system. (c) Real-space fluctuation $\sigma^2_{f_z}(L)$, and (d) structure factor $S_{z=3,4,5,6}(q)$ for each component. Power-law scalings are indicated by solid lines as visual guides.  Here the system size is $N=102400$ and curves in (a,c,d) are shifted for better visualization.}
\label{fig4}
\end{figure}

We further investigate how this universal distribution emerges as jamming transition is approached. Since the average coordination number and each component are linked by definition $Z=\sum zf_z$, one might expect all $f_z$ components to approach limiting values following the same scaling as $\Delta Z$. As shown in Figs.~\ref{fig3}(c)-\ref{fig3}(f), this is true for $z\ne 4$, which exhibit the jamming scaling for fully disordered packings and the characteristic crossovers for near-crystalline packings. However, the fraction of isostatic particles, i.e., $f_4$ ($z=Z_c=4$ in 2D), behaves differently: $f_4$ approaches its plateau value faster and remains constant over a wide pressure range. This is likely because $z=4$ particles naturally satisfy the global Maxwell criterion, while particles with $z\ne 4$ require a more delicate balance as isostaticity is approached.

{\it Hyperuniformity of contacts and density.} -- It has recently been unveiled numerically and proved theoretically that the spatial fluctuation of coordination number is strongly suppressed near the jamming transition, leading to contact hyperuniformity~\cite{hexner2018two,hexner2019can,shang2025jamming}. This is akin to density hyperuniformity, which has also been proposed to be an intrinsic feature of randomly jammed solids \cite{torquato2018hyperuniform,zachary2011hyperuniform}. Here, we examine if contact and density hyperuniformity are universal for marginally jammed solids across the order-disorder spectrum. We quantify contact hyperuniformity by the fluctuation of average coordination number within subsystems of size $L^d$, $\sigma^2_Z(L)$, and the coordination structure factor $S_Z(q)$ ~\cite{hexner2018two}. In addition, we characterize the fluctuation of the coordination-number distribution, $\sigma^2_{f_z}(L)$, and the component structure factor $S_{z=3,4,5,6}(q)$. Considering the particle size dispersity, we quantify the density fluctuation by the spectral density $\chi_{V}(q)$ \cite{torquato2018hyperuniform}. Detailed definitions are given in End Matter.

Figure~\ref{fig4} compares a polydisperse system ($\eta=10^{-2}$) and a quenched binary system near jamming, representative for solids from near-crystalline to fully disordered.
We first consider contact hyperuniformity. Both systems exhibit $\sigma^2_Z(L)\sim L^{-1}$ at large $L$ and $S_Z(q)\sim q^{1.5}$ as $q \to 0$ in Fig.~\ref{fig4}(a). This confirms universal contact hyperuniformity in marginally jammed solids, independent of particle-level structure. The component analysis in Figs.~\ref{fig4}(c) and \ref{fig4}(d) further confirms this generality. In particular, we find $\sigma^2_{f_z} \sim L^{-\nu}$ with $0<\nu<1$ for $z\ne 4$ and $\nu=0$ for $z=4$ in Fig.~\ref{fig4}(c), and correspondingly, $S_{z=3,4,5,6}(q)\sim q^\mu$ with $\mu \approx \nu$ for all components in Fig.~\ref{fig4}(d). This indicates that the fluctuations of coordination-number distribution are suppressed to different extent, being weakly hyperuniform for $z\ne 4$ but fully random for $z=4$. Thus, the class I hyperuniformity of all coordination numbers is not contributed by the same level of hyperuniformity of individual components, but rather a delicate balance between them. This is illustrated by detailed analyses of cross correlations between different coordination-number components in Supplemental Material \cite{supplementary}.

We then examine density hyperuniformity. As shown in Fig.~\ref{fig4}(b), the spectral density of the quenched binary system shows a power-law scaling $\chi_{V}(q)\sim q^{2/3}$ over an intermediate range of $q$, suggesting effective hyperuniformity at corresponding length scales. However, $\chi_{V}(q)$ saturates or weakly upturns as $q$ further decreases, indicating finite density fluctuations at large scales. This lack of true hyperuniformity is consistent with previous studies, often attributed to the sensitivity of density hyperuniformity to rattlers and non-equilibrium protocols for packing generation \cite{torquato2018hyperuniform,ikeda2017large,wu2015search}. The absence of density hyperuniformity is even more evident in near-crystalline packings. Our results thus suggest that, although conjectured to be a unique signature of maximally random jammed packings, density hyperuniformity is not a generic feature of marginally jammed solids and is not necessarily linked to jamming criticality across the spectrum of order and disorder.

{\it Conclusions and discussions.} -- By continuously tuning the structural order of jammed packings from near-crystalline to fully disordered, we unveil universal mechanical and structural characteristics of marginally jammed solids. Our key finding is that jamming criticality universally dominates the mechanics below a critical pressure $p_j$ for any finite disorder, demonstrating that no finite-disorder threshold separates crystal from jamming behaviors. The jamming point (or ``J-line'') can thus approach $\phi_{\rm cp}$, significantly exceeding the upper bound predicted by MF theories related to ideal glasses. This indicates a fundamental decoupling between jamming criticality and glass transition physics (not only that jamming transition is decoupled with glass transition), with the former driven by contact-level disorder and the latter more related to particle-level structure.

We identify a universal coordination-number distribution and contact hyperuniformity in marginally jammed packings regardless of particle-level geometry. These universal signatures in the mechanical network provide strong evidence of general self-organization principles for emergent rigidity upon jamming in physical dimensions. Such nontrivial constraints, absent in mean-field and lattice models, likely cause the observed deviations from mean-field predictions (e.g., $\delta=2.5$ in 2D), as well as the distinction between jamming transition and standard rigidity percolation. Our work thus suggests a fundamental shift of focus for defining amorphous solids from particle-level structure to contact-level mechanical self-organization.


An important next step is to elucidate the precise origin of the universal coordination-number distribution. This is expected to be a central ingredient of a microscopic theory for emergent rigidity in jammed solids. How the incorporation of such steric constraints into MF and effective medium theories would improve the description of jamming criticality in physical dimensions are intriguing problems to explore. Meanwhile, it is interesting to check how these universal characteristics evolve when the requirement of overall homogeneity is lifted, by mixing particles with significantly different sizes or patches with distinct geometries.
To establish a more general understanding of amorphous solids, it is also essential to go beyond simple athermal packings of frictionless disks or spheres, and clarify the role of complex particle shapes, interactions and frictions, as well as thermal fluctuations at finite temperatures. We hope that our findings will initiate further studies in this direction.


{\it Acknowledgments.} -- This work is supported by the National Natural Science Foundation of China (Grant Nos. 12274392 and 12334009). J. Z. also acknowledges support from the China Postdoctoral Science Foundation (Grant No. 2022M713061). We thank the Supercomputing Center of University of Science and Technology of China for the computer time.


\begin{thebibliography}{61}%
\makeatletter
\providecommand \@ifxundefined [1]{%
 \@ifx{#1\undefined}
}%
\providecommand \@ifnum [1]{%
 \ifnum #1\expandafter \@firstoftwo
 \else \expandafter \@secondoftwo
 \fi
}%
\providecommand \@ifx [1]{%
 \ifx #1\expandafter \@firstoftwo
 \else \expandafter \@secondoftwo
 \fi
}%
\providecommand \natexlab [1]{#1}%
\providecommand \enquote  [1]{``#1''}%
\providecommand \bibnamefont  [1]{#1}%
\providecommand \bibfnamefont [1]{#1}%
\providecommand \citenamefont [1]{#1}%
\providecommand \href@noop [0]{\@secondoftwo}%
\providecommand \href [0]{\begingroup \@sanitize@url \@href}%
\providecommand \@href[1]{\@@startlink{#1}\@@href}%
\providecommand \@@href[1]{\endgroup#1\@@endlink}%
\providecommand \@sanitize@url [0]{\catcode `\\12\catcode `\$12\catcode
  `\&12\catcode `\#12\catcode `\^12\catcode `\_12\catcode `\%12\relax}%
\providecommand \@@startlink[1]{}%
\providecommand \@@endlink[0]{}%
\providecommand \url  [0]{\begingroup\@sanitize@url \@url }%
\providecommand \@url [1]{\endgroup\@href {#1}{\urlprefix }}%
\providecommand \urlprefix  [0]{URL }%
\providecommand \Eprint [0]{\href }%
\providecommand \doibase [0]{https://doi.org/}%
\providecommand \selectlanguage [0]{\@gobble}%
\providecommand \bibinfo  [0]{\@secondoftwo}%
\providecommand \bibfield  [0]{\@secondoftwo}%
\providecommand \translation [1]{[#1]}%
\providecommand \BibitemOpen [0]{}%
\providecommand \bibitemStop [0]{}%
\providecommand \bibitemNoStop [0]{.\EOS\space}%
\providecommand \EOS [0]{\spacefactor3000\relax}%
\providecommand \BibitemShut  [1]{\csname bibitem#1\endcsname}%
\let\auto@bib@innerbib\@empty
\bibitem [{\citenamefont {Ashcroft}\ and\ \citenamefont
  {Mermin}(1976)}]{ashcroft1976solid}%
  \BibitemOpen
  \bibfield  {author} {\bibinfo {author} {\bibfnamefont {N.~W.}\ \bibnamefont
  {Ashcroft}}\ and\ \bibinfo {author} {\bibfnamefont {N.~D.}\ \bibnamefont
  {Mermin}},\ }\href@noop {} {\emph {\bibinfo {title} {Solid State Physics}}}\
  (\bibinfo  {publisher} {Thomson Brooks/Cole, Belmont, MA},\ \bibinfo {year}
  {1976})\BibitemShut {NoStop}%
\bibitem [{\citenamefont {Binder}\ and\ \citenamefont
  {Kob}(2011)}]{binder2011glassy}%
  \BibitemOpen
  \bibfield  {author} {\bibinfo {author} {\bibfnamefont {K.}~\bibnamefont
  {Binder}}\ and\ \bibinfo {author} {\bibfnamefont {W.}~\bibnamefont {Kob}},\
  }\href@noop {} {\emph {\bibinfo {title} {Glassy Materials and Disordered
  Solids: An Introduction to Their Statistical Mechanics}}}\ (\bibinfo
  {publisher} {World Scientific, Singapore},\ \bibinfo {year}
  {2011})\BibitemShut {NoStop}%
\bibitem [{\citenamefont {Phillips}(1981)}]{phillips1981amorphous}%
  \BibitemOpen
  \bibinfo {editor} {\bibfnamefont {W.~A.}\ \bibnamefont {Phillips}},\ ed.,\
  \href@noop {} {\emph {\bibinfo {title} {Amorphous Solids: Low-Temperature
  Properties}}}\ (\bibinfo  {publisher} {Springer},\ \bibinfo {year}
  {1981})\BibitemShut {NoStop}%
\bibitem [{\citenamefont {Liu}\ and\ \citenamefont
  {Nagel}(2010)}]{liu2010jamming}%
  \BibitemOpen
  \bibfield  {author} {\bibinfo {author} {\bibfnamefont {A.~J.}\ \bibnamefont
  {Liu}}\ and\ \bibinfo {author} {\bibfnamefont {S.~R.}\ \bibnamefont
  {Nagel}},\ }\bibfield  {title} {\bibinfo {title} {The jamming transition and
  the marginally jammed solid},\ }\href@noop {} {\bibfield  {journal} {\bibinfo
   {journal} {Annu. Rev. Condens. Matter Phys.}\ }\textbf {\bibinfo {volume}
  {1}},\ \bibinfo {pages} {347} (\bibinfo {year} {2010})}\BibitemShut {NoStop}%
\bibitem [{\citenamefont {van Hecke}(2010)}]{van2009jamming}%
  \BibitemOpen
  \bibfield  {author} {\bibinfo {author} {\bibfnamefont {M.}~\bibnamefont {van
  Hecke}},\ }\bibfield  {title} {\bibinfo {title} {Jamming of soft particles:
  geometry, mechanics, scaling and isostaticity},\ }\href@noop {} {\bibfield
  {journal} {\bibinfo  {journal} {J. Phys. Condens. Matter}\ }\textbf {\bibinfo
  {volume} {22}},\ \bibinfo {pages} {033101} (\bibinfo {year}
  {2010})}\BibitemShut {NoStop}%
\bibitem [{\citenamefont {Torquato}\ and\ \citenamefont
  {Stillinger}(2010)}]{torquato2010jammed}%
  \BibitemOpen
  \bibfield  {author} {\bibinfo {author} {\bibfnamefont {S.}~\bibnamefont
  {Torquato}}\ and\ \bibinfo {author} {\bibfnamefont {F.~H.}\ \bibnamefont
  {Stillinger}},\ }\bibfield  {title} {\bibinfo {title} {Jammed hard-particle
  packings: From kepler to bernal and beyond},\ }\href@noop {} {\bibfield
  {journal} {\bibinfo  {journal} {Rev. Mod. Phys.}\ }\textbf {\bibinfo {volume}
  {82}},\ \bibinfo {pages} {2633} (\bibinfo {year} {2010})}\BibitemShut
  {NoStop}%
\bibitem [{\citenamefont {Bi}\ \emph {et~al.}(2015)\citenamefont {Bi},
  \citenamefont {Henkes}, \citenamefont {Daniels},\ and\ \citenamefont
  {Chakraborty}}]{bi2015statistical}%
  \BibitemOpen
  \bibfield  {author} {\bibinfo {author} {\bibfnamefont {D.}~\bibnamefont
  {Bi}}, \bibinfo {author} {\bibfnamefont {S.}~\bibnamefont {Henkes}}, \bibinfo
  {author} {\bibfnamefont {K.~E.}\ \bibnamefont {Daniels}},\ and\ \bibinfo
  {author} {\bibfnamefont {B.}~\bibnamefont {Chakraborty}},\ }\bibfield
  {title} {\bibinfo {title} {The statistical physics of athermal materials},\
  }\href@noop {} {\bibfield  {journal} {\bibinfo  {journal} {Annu. Rev.
  Condens. Matter Phys.}\ }\textbf {\bibinfo {volume} {6}},\ \bibinfo {pages}
  {63} (\bibinfo {year} {2015})}\BibitemShut {NoStop}%
\bibitem [{\citenamefont {Goodrich}\ \emph {et~al.}(2014)\citenamefont
  {Goodrich}, \citenamefont {Liu},\ and\ \citenamefont
  {Nagel}}]{goodrich2014solids}%
  \BibitemOpen
  \bibfield  {author} {\bibinfo {author} {\bibfnamefont {C.~P.}\ \bibnamefont
  {Goodrich}}, \bibinfo {author} {\bibfnamefont {A.~J.}\ \bibnamefont {Liu}},\
  and\ \bibinfo {author} {\bibfnamefont {S.~R.}\ \bibnamefont {Nagel}},\
  }\bibfield  {title} {\bibinfo {title} {Solids between the mechanical extremes
  of order and disorder},\ }\href@noop {} {\bibfield  {journal} {\bibinfo
  {journal} {Nat. Phys.}\ }\textbf {\bibinfo {volume} {10}},\ \bibinfo {pages}
  {578} (\bibinfo {year} {2014})}\BibitemShut {NoStop}%
\bibitem [{\citenamefont {Tong}\ \emph {et~al.}(2015)\citenamefont {Tong},
  \citenamefont {Tan},\ and\ \citenamefont {Xu}}]{tong2015crystals}%
  \BibitemOpen
  \bibfield  {author} {\bibinfo {author} {\bibfnamefont {H.}~\bibnamefont
  {Tong}}, \bibinfo {author} {\bibfnamefont {P.}~\bibnamefont {Tan}},\ and\
  \bibinfo {author} {\bibfnamefont {N.}~\bibnamefont {Xu}},\ }\bibfield
  {title} {\bibinfo {title} {From crystals to disordered crystals: A hidden
  order-disorder transition},\ }\href@noop {} {\bibfield  {journal} {\bibinfo
  {journal} {Sci. Rep.}\ }\textbf {\bibinfo {volume} {5}},\ \bibinfo {pages}
  {15378} (\bibinfo {year} {2015})}\BibitemShut {NoStop}%
\bibitem [{\citenamefont {Charbonneau}\ \emph {et~al.}(2019)\citenamefont
  {Charbonneau}, \citenamefont {Corwin}, \citenamefont {Fu}, \citenamefont
  {Tsekenis},\ and\ \citenamefont {van Der~Naald}}]{charbonneau2019glassy}%
  \BibitemOpen
  \bibfield  {author} {\bibinfo {author} {\bibfnamefont {P.}~\bibnamefont
  {Charbonneau}}, \bibinfo {author} {\bibfnamefont {E.~I.}\ \bibnamefont
  {Corwin}}, \bibinfo {author} {\bibfnamefont {L.}~\bibnamefont {Fu}}, \bibinfo
  {author} {\bibfnamefont {G.}~\bibnamefont {Tsekenis}},\ and\ \bibinfo
  {author} {\bibfnamefont {M.}~\bibnamefont {van Der~Naald}},\ }\bibfield
  {title} {\bibinfo {title} {Glassy, gardner-like phenomenology in minimally
  polydisperse crystalline systems},\ }\href@noop {} {\bibfield  {journal}
  {\bibinfo  {journal} {Phys. Rev. E}\ }\textbf {\bibinfo {volume} {99}},\
  \bibinfo {pages} {020901} (\bibinfo {year} {2019})}\BibitemShut {NoStop}%
\bibitem [{\citenamefont {Tsekenis}(2021)}]{tsekenis2021jamming}%
  \BibitemOpen
  \bibfield  {author} {\bibinfo {author} {\bibfnamefont {G.}~\bibnamefont
  {Tsekenis}},\ }\bibfield  {title} {\bibinfo {title} {Jamming criticality of
  near-crystals},\ }\href@noop {} {\bibfield  {journal} {\bibinfo  {journal}
  {Europhys. Lett.}\ }\textbf {\bibinfo {volume} {135}},\ \bibinfo {pages}
  {36001} (\bibinfo {year} {2021})}\BibitemShut {NoStop}%
\bibitem [{\citenamefont {Ikeda}(2020)}]{ikeda2020jamming}%
  \BibitemOpen
  \bibfield  {author} {\bibinfo {author} {\bibfnamefont {H.}~\bibnamefont
  {Ikeda}},\ }\bibfield  {title} {\bibinfo {title} {Jamming and replica
  symmetry breaking of weakly disordered crystals},\ }\href@noop {} {\bibfield
  {journal} {\bibinfo  {journal} {Phys. Rev. Res.}\ }\textbf {\bibinfo {volume}
  {2}},\ \bibinfo {pages} {033220} (\bibinfo {year} {2020})}\BibitemShut
  {NoStop}%
\bibitem [{\citenamefont {Zhang}\ \emph {et~al.}(2009)\citenamefont {Zhang},
  \citenamefont {Xu}, \citenamefont {Chen}, \citenamefont {Yunker},
  \citenamefont {Alsayed}, \citenamefont {Aptowicz}, \citenamefont {Habdas},
  \citenamefont {Liu}, \citenamefont {Nagel},\ and\ \citenamefont
  {Yodh}}]{zhang2009thermal}%
  \BibitemOpen
  \bibfield  {author} {\bibinfo {author} {\bibfnamefont {Z.}~\bibnamefont
  {Zhang}}, \bibinfo {author} {\bibfnamefont {N.}~\bibnamefont {Xu}}, \bibinfo
  {author} {\bibfnamefont {D.~T.~N.}\ \bibnamefont {Chen}}, \bibinfo {author}
  {\bibfnamefont {P.}~\bibnamefont {Yunker}}, \bibinfo {author} {\bibfnamefont
  {A.~M.}\ \bibnamefont {Alsayed}}, \bibinfo {author} {\bibfnamefont {K.~B.}\
  \bibnamefont {Aptowicz}}, \bibinfo {author} {\bibfnamefont {P.}~\bibnamefont
  {Habdas}}, \bibinfo {author} {\bibfnamefont {A.~J.}\ \bibnamefont {Liu}},
  \bibinfo {author} {\bibfnamefont {S.~R.}\ \bibnamefont {Nagel}},\ and\
  \bibinfo {author} {\bibfnamefont {A.~G.}\ \bibnamefont {Yodh}},\ }\bibfield
  {title} {\bibinfo {title} {Thermal vestige of the zero-temperature jamming
  transition},\ }\href@noop {} {\bibfield  {journal} {\bibinfo  {journal}
  {Nature}\ }\textbf {\bibinfo {volume} {459}},\ \bibinfo {pages} {230}
  (\bibinfo {year} {2009})}\BibitemShut {NoStop}%
\bibitem [{\citenamefont {Degiuli}\ \emph {et~al.}(2015)\citenamefont
  {Degiuli}, \citenamefont {Lerner},\ and\ \citenamefont
  {Wyart}}]{degiuli2015theory}%
  \BibitemOpen
  \bibfield  {author} {\bibinfo {author} {\bibfnamefont {E.}~\bibnamefont
  {Degiuli}}, \bibinfo {author} {\bibfnamefont {E.}~\bibnamefont {Lerner}},\
  and\ \bibinfo {author} {\bibfnamefont {M.}~\bibnamefont {Wyart}},\ }\bibfield
   {title} {\bibinfo {title} {Theory of the jamming transition at finite
  temperature},\ }\href@noop {} {\bibfield  {journal} {\bibinfo  {journal} {J.
  Chem. Phys.}\ }\textbf {\bibinfo {volume} {142}},\ \bibinfo {pages} {164503}
  (\bibinfo {year} {2015})}\BibitemShut {NoStop}%
\bibitem [{\citenamefont {Ikeda}\ \emph {et~al.}(2013)\citenamefont {Ikeda},
  \citenamefont {Berthier},\ and\ \citenamefont {Biroli}}]{ikeda2013dynamic}%
  \BibitemOpen
  \bibfield  {author} {\bibinfo {author} {\bibfnamefont {A.}~\bibnamefont
  {Ikeda}}, \bibinfo {author} {\bibfnamefont {L.}~\bibnamefont {Berthier}},\
  and\ \bibinfo {author} {\bibfnamefont {G.}~\bibnamefont {Biroli}},\
  }\bibfield  {title} {\bibinfo {title} {Dynamic criticality at the jamming
  transition},\ }\href@noop {} {\bibfield  {journal} {\bibinfo  {journal} {J.
  Chem. Phys.}\ }\textbf {\bibinfo {volume} {138}},\ \bibinfo {pages} {12A507}
  (\bibinfo {year} {2013})}\BibitemShut {NoStop}%
\bibitem [{\citenamefont {Heussinger}\ and\ \citenamefont
  {Barrat}(2009)}]{heussinger2009jamming}%
  \BibitemOpen
  \bibfield  {author} {\bibinfo {author} {\bibfnamefont {C.}~\bibnamefont
  {Heussinger}}\ and\ \bibinfo {author} {\bibfnamefont {J.-L.}\ \bibnamefont
  {Barrat}},\ }\bibfield  {title} {\bibinfo {title} {Jamming transition as
  probed by quasistatic shear flow},\ }\href@noop {} {\bibfield  {journal}
  {\bibinfo  {journal} {Phys. Rev. Lett.}\ }\textbf {\bibinfo {volume} {102}},\
  \bibinfo {pages} {218303} (\bibinfo {year} {2009})}\BibitemShut {NoStop}%
\bibitem [{\citenamefont {Olsson}(2019)}]{olsson2019dimensionality}%
  \BibitemOpen
  \bibfield  {author} {\bibinfo {author} {\bibfnamefont {P.}~\bibnamefont
  {Olsson}},\ }\bibfield  {title} {\bibinfo {title} {Dimensionality and
  viscosity exponent in shear-driven jamming},\ }\href@noop {} {\bibfield
  {journal} {\bibinfo  {journal} {Phys. Rev. Lett.}\ }\textbf {\bibinfo
  {volume} {122}},\ \bibinfo {pages} {108003} (\bibinfo {year}
  {2019})}\BibitemShut {NoStop}%
\bibitem [{\citenamefont {Peshkov}\ and\ \citenamefont
  {Teitel}(2021)}]{peshkov2021critical}%
  \BibitemOpen
  \bibfield  {author} {\bibinfo {author} {\bibfnamefont {A.}~\bibnamefont
  {Peshkov}}\ and\ \bibinfo {author} {\bibfnamefont {S.}~\bibnamefont
  {Teitel}},\ }\bibfield  {title} {\bibinfo {title} {Critical scaling of
  compression-driven jamming of athermal frictionless spheres in suspension},\
  }\href@noop {} {\bibfield  {journal} {\bibinfo  {journal} {Phys. Rev. E}\
  }\textbf {\bibinfo {volume} {103}},\ \bibinfo {pages} {L040901} (\bibinfo
  {year} {2021})}\BibitemShut {NoStop}%
\bibitem [{\citenamefont {Peshkov}\ and\ \citenamefont
  {Teitel}(2022)}]{peshkov2022universality}%
  \BibitemOpen
  \bibfield  {author} {\bibinfo {author} {\bibfnamefont {A.}~\bibnamefont
  {Peshkov}}\ and\ \bibinfo {author} {\bibfnamefont {S.}~\bibnamefont
  {Teitel}},\ }\bibfield  {title} {\bibinfo {title} {Universality of
  stress-anisotropic and stress-isotropic jamming of frictionless spheres in
  three dimensions: Uniaxial versus isotropic compression},\ }\href@noop {}
  {\bibfield  {journal} {\bibinfo  {journal} {Phys. Rev. E}\ }\textbf {\bibinfo
  {volume} {105}},\ \bibinfo {pages} {024902} (\bibinfo {year}
  {2022})}\BibitemShut {NoStop}%
\bibitem [{\citenamefont {Ellenbroek}\ \emph {et~al.}(2009)\citenamefont
  {Ellenbroek}, \citenamefont {Zeravcic}, \citenamefont {van Saarloos},\ and\
  \citenamefont {van Hecke}}]{ellenbroek2009non}%
  \BibitemOpen
  \bibfield  {author} {\bibinfo {author} {\bibfnamefont {W.~G.}\ \bibnamefont
  {Ellenbroek}}, \bibinfo {author} {\bibfnamefont {Z.}~\bibnamefont
  {Zeravcic}}, \bibinfo {author} {\bibfnamefont {W.}~\bibnamefont {van
  Saarloos}},\ and\ \bibinfo {author} {\bibfnamefont {M.}~\bibnamefont {van
  Hecke}},\ }\bibfield  {title} {\bibinfo {title} {Non-affine response: Jammed
  packings vs. spring networks},\ }\href@noop {} {\bibfield  {journal}
  {\bibinfo  {journal} {Europhys. Lett.}\ }\textbf {\bibinfo {volume} {87}},\
  \bibinfo {pages} {34004} (\bibinfo {year} {2009})}\BibitemShut {NoStop}%
\bibitem [{\citenamefont {Ellenbroek}\ \emph {et~al.}(2015)\citenamefont
  {Ellenbroek}, \citenamefont {Hagh}, \citenamefont {Kumar}, \citenamefont
  {Thorpe},\ and\ \citenamefont {Van~Hecke}}]{ellenbroek2015rigidity}%
  \BibitemOpen
  \bibfield  {author} {\bibinfo {author} {\bibfnamefont {W.~G.}\ \bibnamefont
  {Ellenbroek}}, \bibinfo {author} {\bibfnamefont {V.~F.}\ \bibnamefont
  {Hagh}}, \bibinfo {author} {\bibfnamefont {A.}~\bibnamefont {Kumar}},
  \bibinfo {author} {\bibfnamefont {M.}~\bibnamefont {Thorpe}},\ and\ \bibinfo
  {author} {\bibfnamefont {M.}~\bibnamefont {Van~Hecke}},\ }\bibfield  {title}
  {\bibinfo {title} {Rigidity loss in disordered systems: Three scenarios},\
  }\href@noop {} {\bibfield  {journal} {\bibinfo  {journal} {Phys. Rev. Lett.}\
  }\textbf {\bibinfo {volume} {114}},\ \bibinfo {pages} {135501} (\bibinfo
  {year} {2015})}\BibitemShut {NoStop}%
\bibitem [{\citenamefont {{O'Hern}}\ \emph {et~al.}(2003)\citenamefont
  {{O'Hern}}, \citenamefont {Silbert}, \citenamefont {Liu},\ and\ \citenamefont
  {Nagel}}]{o2003jamming}%
  \BibitemOpen
  \bibfield  {author} {\bibinfo {author} {\bibfnamefont {C.~S.}\ \bibnamefont
  {{O'Hern}}}, \bibinfo {author} {\bibfnamefont {L.~E.}\ \bibnamefont
  {Silbert}}, \bibinfo {author} {\bibfnamefont {A.~J.}\ \bibnamefont {Liu}},\
  and\ \bibinfo {author} {\bibfnamefont {S.~R.}\ \bibnamefont {Nagel}},\
  }\bibfield  {title} {\bibinfo {title} {Jamming at zero temperature and zero
  applied stress: The epitome of disorder},\ }\href@noop {} {\bibfield
  {journal} {\bibinfo  {journal} {Phys. Rev. E}\ }\textbf {\bibinfo {volume}
  {68}},\ \bibinfo {pages} {011306} (\bibinfo {year} {2003})}\BibitemShut
  {NoStop}%
\bibitem [{\citenamefont {Feng}\ \emph {et~al.}(1985)\citenamefont {Feng},
  \citenamefont {Thorpe},\ and\ \citenamefont {Garboczi}}]{feng1985effective}%
  \BibitemOpen
  \bibfield  {author} {\bibinfo {author} {\bibfnamefont {S.}~\bibnamefont
  {Feng}}, \bibinfo {author} {\bibfnamefont {M.}~\bibnamefont {Thorpe}},\ and\
  \bibinfo {author} {\bibfnamefont {E.}~\bibnamefont {Garboczi}},\ }\bibfield
  {title} {\bibinfo {title} {Effective-medium theory of percolation on
  central-force elastic networks},\ }\href@noop {} {\bibfield  {journal}
  {\bibinfo  {journal} {Phys. Rev. B}\ }\textbf {\bibinfo {volume} {31}},\
  \bibinfo {pages} {276} (\bibinfo {year} {1985})}\BibitemShut {NoStop}%
\bibitem [{\citenamefont {Jacobs}\ and\ \citenamefont
  {Thorpe}(1995)}]{jacobs1995generic}%
  \BibitemOpen
  \bibfield  {author} {\bibinfo {author} {\bibfnamefont {D.~J.}\ \bibnamefont
  {Jacobs}}\ and\ \bibinfo {author} {\bibfnamefont {M.~F.}\ \bibnamefont
  {Thorpe}},\ }\bibfield  {title} {\bibinfo {title} {Generic rigidity
  percolation: the pebble game},\ }\href@noop {} {\bibfield  {journal}
  {\bibinfo  {journal} {Phys. Rev. Lett.}\ }\textbf {\bibinfo {volume} {75}},\
  \bibinfo {pages} {4051} (\bibinfo {year} {1995})}\BibitemShut {NoStop}%
\bibitem [{\citenamefont {Hagh}\ \emph {et~al.}(2019)\citenamefont {Hagh},
  \citenamefont {Corwin}, \citenamefont {Stephenson},\ and\ \citenamefont
  {Thorpe}}]{hagh2019broader}%
  \BibitemOpen
  \bibfield  {author} {\bibinfo {author} {\bibfnamefont {V.~F.}\ \bibnamefont
  {Hagh}}, \bibinfo {author} {\bibfnamefont {E.~I.}\ \bibnamefont {Corwin}},
  \bibinfo {author} {\bibfnamefont {K.}~\bibnamefont {Stephenson}},\ and\
  \bibinfo {author} {\bibfnamefont {M.}~\bibnamefont {Thorpe}},\ }\bibfield
  {title} {\bibinfo {title} {A broader view on jamming: from spring networks to
  circle packings},\ }\href@noop {} {\bibfield  {journal} {\bibinfo  {journal}
  {Soft matter}\ }\textbf {\bibinfo {volume} {15}},\ \bibinfo {pages} {3076}
  (\bibinfo {year} {2019})}\BibitemShut {NoStop}%
\bibitem [{\citenamefont {Liarte}\ \emph {et~al.}(2019)\citenamefont {Liarte},
  \citenamefont {Mao}, \citenamefont {Stenull},\ and\ \citenamefont
  {Lubensky}}]{liarte2019jamming}%
  \BibitemOpen
  \bibfield  {author} {\bibinfo {author} {\bibfnamefont {D.~B.}\ \bibnamefont
  {Liarte}}, \bibinfo {author} {\bibfnamefont {X.}~\bibnamefont {Mao}},
  \bibinfo {author} {\bibfnamefont {O.}~\bibnamefont {Stenull}},\ and\ \bibinfo
  {author} {\bibfnamefont {T.}~\bibnamefont {Lubensky}},\ }\bibfield  {title}
  {\bibinfo {title} {Jamming as a multicritical point},\ }\href@noop {}
  {\bibfield  {journal} {\bibinfo  {journal} {Phys. Rev. Lett.}\ }\textbf
  {\bibinfo {volume} {122}},\ \bibinfo {pages} {128006} (\bibinfo {year}
  {2019})}\BibitemShut {NoStop}%
\bibitem [{\citenamefont {Parisi}\ and\ \citenamefont
  {Zamponi}(2010)}]{parisi2010mean}%
  \BibitemOpen
  \bibfield  {author} {\bibinfo {author} {\bibfnamefont {G.}~\bibnamefont
  {Parisi}}\ and\ \bibinfo {author} {\bibfnamefont {F.}~\bibnamefont
  {Zamponi}},\ }\bibfield  {title} {\bibinfo {title} {Mean-field theory of hard
  sphere glasses and jamming},\ }\href@noop {} {\bibfield  {journal} {\bibinfo
  {journal} {Rev. Mod. Phys.}\ }\textbf {\bibinfo {volume} {82}},\ \bibinfo
  {pages} {789} (\bibinfo {year} {2010})}\BibitemShut {NoStop}%
\bibitem [{\citenamefont {Kurchan}\ \emph {et~al.}(2012)\citenamefont
  {Kurchan}, \citenamefont {Parisi},\ and\ \citenamefont
  {Zamponi}}]{kurchan2012exact}%
  \BibitemOpen
  \bibfield  {author} {\bibinfo {author} {\bibfnamefont {J.}~\bibnamefont
  {Kurchan}}, \bibinfo {author} {\bibfnamefont {G.}~\bibnamefont {Parisi}},\
  and\ \bibinfo {author} {\bibfnamefont {F.}~\bibnamefont {Zamponi}},\
  }\bibfield  {title} {\bibinfo {title} {Exact theory of dense amorphous hard
  spheres in high dimension. {I}. the free energy},\ }\href@noop {} {\bibfield
  {journal} {\bibinfo  {journal} {J. Stat. Mech.}\ }\textbf {\bibinfo {volume}
  {2012}},\ \bibinfo {pages} {P10012} (\bibinfo {year} {2012})}\BibitemShut
  {NoStop}%
\bibitem [{\citenamefont {Charbonneau}\ \emph {et~al.}(2017)\citenamefont
  {Charbonneau}, \citenamefont {Kurchan}, \citenamefont {Parisi}, \citenamefont
  {Urbani},\ and\ \citenamefont {Zamponi}}]{charbonneau2017glass}%
  \BibitemOpen
  \bibfield  {author} {\bibinfo {author} {\bibfnamefont {P.}~\bibnamefont
  {Charbonneau}}, \bibinfo {author} {\bibfnamefont {J.}~\bibnamefont
  {Kurchan}}, \bibinfo {author} {\bibfnamefont {G.}~\bibnamefont {Parisi}},
  \bibinfo {author} {\bibfnamefont {P.}~\bibnamefont {Urbani}},\ and\ \bibinfo
  {author} {\bibfnamefont {F.}~\bibnamefont {Zamponi}},\ }\bibfield  {title}
  {\bibinfo {title} {Glass and jamming transitions: From exact results to
  finite-dimensional descriptions},\ }\href@noop {} {\bibfield  {journal}
  {\bibinfo  {journal} {Annu. Rev. Condens. Matter Phys.}\ }\textbf {\bibinfo
  {volume} {8}},\ \bibinfo {pages} {265} (\bibinfo {year} {2017})}\BibitemShut
  {NoStop}%
\bibitem [{\citenamefont {Charbonneau}\ \emph {et~al.}(2014)\citenamefont
  {Charbonneau}, \citenamefont {Kurchan}, \citenamefont {Parisi}, \citenamefont
  {Urbani},\ and\ \citenamefont {Zamponi}}]{charbonneau2014fractal}%
  \BibitemOpen
  \bibfield  {author} {\bibinfo {author} {\bibfnamefont {P.}~\bibnamefont
  {Charbonneau}}, \bibinfo {author} {\bibfnamefont {J.}~\bibnamefont
  {Kurchan}}, \bibinfo {author} {\bibfnamefont {G.}~\bibnamefont {Parisi}},
  \bibinfo {author} {\bibfnamefont {P.}~\bibnamefont {Urbani}},\ and\ \bibinfo
  {author} {\bibfnamefont {F.}~\bibnamefont {Zamponi}},\ }\bibfield  {title}
  {\bibinfo {title} {Fractal free energy landscapes in structural glasses},\
  }\href@noop {} {\bibfield  {journal} {\bibinfo  {journal} {Nat. Commun.}\
  }\textbf {\bibinfo {volume} {5}},\ \bibinfo {pages} {3725} (\bibinfo {year}
  {2014})}\BibitemShut {NoStop}%
\bibitem [{\citenamefont {Tanaka}(2012)}]{tanaka2012bond}%
  \BibitemOpen
  \bibfield  {author} {\bibinfo {author} {\bibfnamefont {H.}~\bibnamefont
  {Tanaka}},\ }\bibfield  {title} {\bibinfo {title} {Bond orientational order
  in liquids: Towards a unified description of water-like anomalies,
  liquid-liquid transition, glass transition, and crystallization},\
  }\href@noop {} {\bibfield  {journal} {\bibinfo  {journal} {Eur. Phys. J. E}\
  }\textbf {\bibinfo {volume} {35}},\ \bibinfo {pages} {113} (\bibinfo {year}
  {2012})}\BibitemShut {NoStop}%
\bibitem [{\citenamefont {Tanaka}\ \emph {et~al.}(2019)\citenamefont {Tanaka},
  \citenamefont {Tong}, \citenamefont {Shi},\ and\ \citenamefont
  {Russo}}]{tanaka2019revealing}%
  \BibitemOpen
  \bibfield  {author} {\bibinfo {author} {\bibfnamefont {H.}~\bibnamefont
  {Tanaka}}, \bibinfo {author} {\bibfnamefont {H.}~\bibnamefont {Tong}},
  \bibinfo {author} {\bibfnamefont {R.}~\bibnamefont {Shi}},\ and\ \bibinfo
  {author} {\bibfnamefont {J.}~\bibnamefont {Russo}},\ }\bibfield  {title}
  {\bibinfo {title} {Revealing key structural features hidden in liquids and
  glasses},\ }\href@noop {} {\bibfield  {journal} {\bibinfo  {journal} {Nat.
  Rev. Phys.}\ ,\ \bibinfo {pages} {1}} (\bibinfo {year} {2019})}\BibitemShut
  {NoStop}%
\bibitem [{\citenamefont {Wyart}\ \emph {et~al.}(2005)\citenamefont {Wyart},
  \citenamefont {Silbert}, \citenamefont {Nagel},\ and\ \citenamefont
  {Witten}}]{wyart2005effects}%
  \BibitemOpen
  \bibfield  {author} {\bibinfo {author} {\bibfnamefont {M.}~\bibnamefont
  {Wyart}}, \bibinfo {author} {\bibfnamefont {L.~E.}\ \bibnamefont {Silbert}},
  \bibinfo {author} {\bibfnamefont {S.~R.}\ \bibnamefont {Nagel}},\ and\
  \bibinfo {author} {\bibfnamefont {T.~A.}\ \bibnamefont {Witten}},\ }\bibfield
   {title} {\bibinfo {title} {Effects of compression on the vibrational modes
  of marginally jammed solids},\ }\href@noop {} {\bibfield  {journal} {\bibinfo
   {journal} {Phys. Rev. E}\ }\textbf {\bibinfo {volume} {72}},\ \bibinfo
  {pages} {051306} (\bibinfo {year} {2005})}\BibitemShut {NoStop}%
\bibitem [{\citenamefont {Wyart}(2005)}]{wyart2005rigidity}%
  \BibitemOpen
  \bibfield  {author} {\bibinfo {author} {\bibfnamefont {M.}~\bibnamefont
  {Wyart}},\ }\bibfield  {title} {\bibinfo {title} {On the rigidity of
  amorphous solids},\ }\href@noop {} {\bibfield  {journal} {\bibinfo  {journal}
  {Ann. Phys.}\ }\textbf {\bibinfo {volume} {30}},\ \bibinfo {pages} {1}
  (\bibinfo {year} {2005})}\BibitemShut {NoStop}%
\bibitem [{\citenamefont {Wyart}(2010)}]{wyart2010scaling}%
  \BibitemOpen
  \bibfield  {author} {\bibinfo {author} {\bibfnamefont {M.}~\bibnamefont
  {Wyart}},\ }\bibfield  {title} {\bibinfo {title} {Scaling of phononic
  transport with connectivity in amorphous solids},\ }\href@noop {} {\bibfield
  {journal} {\bibinfo  {journal} {Europhys. Lett.}\ }\textbf {\bibinfo {volume}
  {89}},\ \bibinfo {pages} {64001} (\bibinfo {year} {2010})}\BibitemShut
  {NoStop}%
\bibitem [{\citenamefont {Mao}\ \emph {et~al.}(2010)\citenamefont {Mao},
  \citenamefont {Xu},\ and\ \citenamefont {Lubensky}}]{mao2010soft}%
  \BibitemOpen
  \bibfield  {author} {\bibinfo {author} {\bibfnamefont {X.}~\bibnamefont
  {Mao}}, \bibinfo {author} {\bibfnamefont {N.}~\bibnamefont {Xu}},\ and\
  \bibinfo {author} {\bibfnamefont {T.}~\bibnamefont {Lubensky}},\ }\bibfield
  {title} {\bibinfo {title} {Soft modes and elasticity of nearly isostatic
  lattices: Randomness and dissipation},\ }\href@noop {} {\bibfield  {journal}
  {\bibinfo  {journal} {Phys. Rev. Lett.}\ }\textbf {\bibinfo {volume} {104}},\
  \bibinfo {pages} {085504} (\bibinfo {year} {2010})}\BibitemShut {NoStop}%
\bibitem [{\citenamefont {DeGiuli}\ \emph {et~al.}(2014)\citenamefont
  {DeGiuli}, \citenamefont {Laversanne-Finot}, \citenamefont {D{\"u}ring},
  \citenamefont {Lerner},\ and\ \citenamefont {Wyart}}]{degiuli2014effects}%
  \BibitemOpen
  \bibfield  {author} {\bibinfo {author} {\bibfnamefont {E.}~\bibnamefont
  {DeGiuli}}, \bibinfo {author} {\bibfnamefont {A.}~\bibnamefont
  {Laversanne-Finot}}, \bibinfo {author} {\bibfnamefont {G.}~\bibnamefont
  {D{\"u}ring}}, \bibinfo {author} {\bibfnamefont {E.}~\bibnamefont {Lerner}},\
  and\ \bibinfo {author} {\bibfnamefont {M.}~\bibnamefont {Wyart}},\ }\bibfield
   {title} {\bibinfo {title} {Effects of coordination and pressure on sound
  attenuation, boson peak and elasticity in amorphous solids},\ }\href@noop {}
  {\bibfield  {journal} {\bibinfo  {journal} {Soft Matter}\ }\textbf {\bibinfo
  {volume} {10}},\ \bibinfo {pages} {5628} (\bibinfo {year}
  {2014})}\BibitemShut {NoStop}%
\bibitem [{\citenamefont {Makse}\ \emph {et~al.}(1999)\citenamefont {Makse},
  \citenamefont {Gland}, \citenamefont {Johnson},\ and\ \citenamefont
  {Schwartz}}]{makse1999effective}%
  \BibitemOpen
  \bibfield  {author} {\bibinfo {author} {\bibfnamefont {H.~A.}\ \bibnamefont
  {Makse}}, \bibinfo {author} {\bibfnamefont {N.}~\bibnamefont {Gland}},
  \bibinfo {author} {\bibfnamefont {D.~L.}\ \bibnamefont {Johnson}},\ and\
  \bibinfo {author} {\bibfnamefont {L.~M.}\ \bibnamefont {Schwartz}},\
  }\bibfield  {title} {\bibinfo {title} {Why effective medium theory fails in
  granular materials},\ }\href@noop {} {\bibfield  {journal} {\bibinfo
  {journal} {Phys. Rev. Lett.}\ }\textbf {\bibinfo {volume} {83}},\ \bibinfo
  {pages} {5070} (\bibinfo {year} {1999})}\BibitemShut {NoStop}%
\bibitem [{\citenamefont {Mizuno}\ \emph {et~al.}(2013)\citenamefont {Mizuno},
  \citenamefont {Mossa},\ and\ \citenamefont {Barrat}}]{mizuno2013elastic}%
  \BibitemOpen
  \bibfield  {author} {\bibinfo {author} {\bibfnamefont {H.}~\bibnamefont
  {Mizuno}}, \bibinfo {author} {\bibfnamefont {S.}~\bibnamefont {Mossa}},\ and\
  \bibinfo {author} {\bibfnamefont {J.-L.}\ \bibnamefont {Barrat}},\ }\bibfield
   {title} {\bibinfo {title} {Elastic heterogeneity, vibrational states, and
  thermal conductivity across an amorphisation transition},\ }\href@noop {}
  {\bibfield  {journal} {\bibinfo  {journal} {Europhys. Lett.}\ }\textbf
  {\bibinfo {volume} {104}},\ \bibinfo {pages} {56001} (\bibinfo {year}
  {2013})}\BibitemShut {NoStop}%
\bibitem [{\citenamefont {Mizuno}\ \emph {et~al.}(2014)\citenamefont {Mizuno},
  \citenamefont {Mossa},\ and\ \citenamefont {Barrat}}]{mizuno2014acoustic}%
  \BibitemOpen
  \bibfield  {author} {\bibinfo {author} {\bibfnamefont {H.}~\bibnamefont
  {Mizuno}}, \bibinfo {author} {\bibfnamefont {S.}~\bibnamefont {Mossa}},\ and\
  \bibinfo {author} {\bibfnamefont {J.-L.}\ \bibnamefont {Barrat}},\ }\bibfield
   {title} {\bibinfo {title} {Acoustic excitations and elastic heterogeneities
  in disordered solids},\ }\href@noop {} {\bibfield  {journal} {\bibinfo
  {journal} {Proc. Natl Acad. Sci. USA}\ }\textbf {\bibinfo {volume} {111}},\
  \bibinfo {pages} {11949} (\bibinfo {year} {2014})}\BibitemShut {NoStop}%
\bibitem [{\citenamefont {Gibbs}(1878)}]{gibbs1878equilibrium}%
  \BibitemOpen
  \bibfield  {author} {\bibinfo {author} {\bibfnamefont {J.~W.}\ \bibnamefont
  {Gibbs}},\ }\bibfield  {title} {\bibinfo {title} {On the equilibrium of
  heterogeneous substances},\ }\href@noop {} {\bibfield  {journal} {\bibinfo
  {journal} {Trans. Conn. Acad. Arts Sci.}\ }\textbf {\bibinfo {volume} {3}},\
  \bibinfo {pages} {108–248} (\bibinfo {year} {1878})}\BibitemShut {NoStop}%
\bibitem [{\citenamefont {Bitzek}\ \emph {et~al.}(2006)\citenamefont {Bitzek},
  \citenamefont {Koskinen}, \citenamefont {G{\"a}hler}, \citenamefont
  {Moseler},\ and\ \citenamefont {Gumbsch}}]{bitzek2006structural}%
  \BibitemOpen
  \bibfield  {author} {\bibinfo {author} {\bibfnamefont {E.}~\bibnamefont
  {Bitzek}}, \bibinfo {author} {\bibfnamefont {P.}~\bibnamefont {Koskinen}},
  \bibinfo {author} {\bibfnamefont {F.}~\bibnamefont {G{\"a}hler}}, \bibinfo
  {author} {\bibfnamefont {M.}~\bibnamefont {Moseler}},\ and\ \bibinfo {author}
  {\bibfnamefont {P.}~\bibnamefont {Gumbsch}},\ }\bibfield  {title} {\bibinfo
  {title} {Structural relaxation made simple},\ }\href@noop {} {\bibfield
  {journal} {\bibinfo  {journal} {Phys. Rev. Lett.}\ }\textbf {\bibinfo
  {volume} {97}},\ \bibinfo {pages} {170201} (\bibinfo {year}
  {2006})}\BibitemShut {NoStop}%
\bibitem [{sup()}]{supplementary}%
  \BibitemOpen
  \href@noop {} {\bibinfo  {journal} {See Supplemental Material at
  http://link.aps.org/supplemental/XXX for further characterizations and
  discussions}\ }\BibitemShut {NoStop}%
\bibitem [{\citenamefont {Goodrich}\ \emph {et~al.}(2012)\citenamefont
  {Goodrich}, \citenamefont {Liu},\ and\ \citenamefont
  {Nagel}}]{goodrich2012finite}%
  \BibitemOpen
\bibfield  {journal} {  }\bibfield  {author} {\bibinfo {author} {\bibfnamefont
  {C.~P.}\ \bibnamefont {Goodrich}}, \bibinfo {author} {\bibfnamefont {A.~J.}\
  \bibnamefont {Liu}},\ and\ \bibinfo {author} {\bibfnamefont {S.~R.}\
  \bibnamefont {Nagel}},\ }\bibfield  {title} {\bibinfo {title} {Finite-size
  scaling at the jamming transition},\ }\href@noop {} {\bibfield  {journal}
  {\bibinfo  {journal} {Phys. Rev. Lett.}\ }\textbf {\bibinfo {volume} {109}},\
  \bibinfo {pages} {095704} (\bibinfo {year} {2012})}\BibitemShut {NoStop}%
\bibitem [{\citenamefont {Hexner}\ \emph {et~al.}(2018)\citenamefont {Hexner},
  \citenamefont {Liu},\ and\ \citenamefont {Nagel}}]{hexner2018two}%
  \BibitemOpen
  \bibfield  {author} {\bibinfo {author} {\bibfnamefont {D.}~\bibnamefont
  {Hexner}}, \bibinfo {author} {\bibfnamefont {A.~J.}\ \bibnamefont {Liu}},\
  and\ \bibinfo {author} {\bibfnamefont {S.~R.}\ \bibnamefont {Nagel}},\
  }\bibfield  {title} {\bibinfo {title} {Two diverging length scales in the
  structure of jammed packings},\ }\href@noop {} {\bibfield  {journal}
  {\bibinfo  {journal} {Phys. Rev. Lett.}\ }\textbf {\bibinfo {volume} {121}},\
  \bibinfo {pages} {115501} (\bibinfo {year} {2018})}\BibitemShut {NoStop}%
\bibitem [{\citenamefont {Chaudhuri}\ \emph {et~al.}(2010)\citenamefont
  {Chaudhuri}, \citenamefont {Berthier},\ and\ \citenamefont
  {Sastry}}]{chaudhuri2010jamming}%
  \BibitemOpen
  \bibfield  {author} {\bibinfo {author} {\bibfnamefont {P.}~\bibnamefont
  {Chaudhuri}}, \bibinfo {author} {\bibfnamefont {L.}~\bibnamefont
  {Berthier}},\ and\ \bibinfo {author} {\bibfnamefont {S.}~\bibnamefont
  {Sastry}},\ }\bibfield  {title} {\bibinfo {title} {Jamming transitions in
  amorphous packings of frictionless spheres occur over a continuous range of
  volume fractions},\ }\href@noop {} {\bibfield  {journal} {\bibinfo  {journal}
  {Phys. Rev. Lett.}\ }\textbf {\bibinfo {volume} {104}},\ \bibinfo {pages}
  {165701} (\bibinfo {year} {2010})}\BibitemShut {NoStop}%
\bibitem [{\citenamefont {Ozawa}\ \emph {et~al.}(2017)\citenamefont {Ozawa},
  \citenamefont {Berthier},\ and\ \citenamefont
  {Coslovich}}]{ozawa2017exploring}%
  \BibitemOpen
  \bibfield  {author} {\bibinfo {author} {\bibfnamefont {M.}~\bibnamefont
  {Ozawa}}, \bibinfo {author} {\bibfnamefont {L.}~\bibnamefont {Berthier}},\
  and\ \bibinfo {author} {\bibfnamefont {D.}~\bibnamefont {Coslovich}},\
  }\bibfield  {title} {\bibinfo {title} {Exploring the jamming transition over
  a wide range of critical densities},\ }\href@noop {} {\bibfield  {journal}
  {\bibinfo  {journal} {SciPost Phys.}\ }\textbf {\bibinfo {volume} {3}},\
  \bibinfo {pages} {027} (\bibinfo {year} {2017})}\BibitemShut {NoStop}%
\bibitem [{\citenamefont {Xing}\ \emph {et~al.}(2024)\citenamefont {Xing},
  \citenamefont {Yuan}, \citenamefont {Yuan}, \citenamefont {Zhang},
  \citenamefont {Zeng}, \citenamefont {Zheng}, \citenamefont {Xia},\ and\
  \citenamefont {Wang}}]{xing2024origin}%
  \BibitemOpen
  \bibfield  {author} {\bibinfo {author} {\bibfnamefont {Y.}~\bibnamefont
  {Xing}}, \bibinfo {author} {\bibfnamefont {Y.}~\bibnamefont {Yuan}}, \bibinfo
  {author} {\bibfnamefont {H.}~\bibnamefont {Yuan}}, \bibinfo {author}
  {\bibfnamefont {S.}~\bibnamefont {Zhang}}, \bibinfo {author} {\bibfnamefont
  {Z.}~\bibnamefont {Zeng}}, \bibinfo {author} {\bibfnamefont {X.}~\bibnamefont
  {Zheng}}, \bibinfo {author} {\bibfnamefont {C.}~\bibnamefont {Xia}},\ and\
  \bibinfo {author} {\bibfnamefont {Y.}~\bibnamefont {Wang}},\ }\bibfield
  {title} {\bibinfo {title} {Origin of the critical state in sheared granular
  materials},\ }\href@noop {} {\bibfield  {journal} {\bibinfo  {journal} {Nat.
  Phys.}\ ,\ \bibinfo {pages} {1}} (\bibinfo {year} {2024})}\BibitemShut
  {NoStop}%
\bibitem [{\citenamefont {Corwin}\ \emph {et~al.}(2005)\citenamefont {Corwin},
  \citenamefont {Jaeger},\ and\ \citenamefont {Nagel}}]{corwin2005structural}%
  \BibitemOpen
  \bibfield  {author} {\bibinfo {author} {\bibfnamefont {E.~I.}\ \bibnamefont
  {Corwin}}, \bibinfo {author} {\bibfnamefont {H.~M.}\ \bibnamefont {Jaeger}},\
  and\ \bibinfo {author} {\bibfnamefont {S.~R.}\ \bibnamefont {Nagel}},\
  }\bibfield  {title} {\bibinfo {title} {Structural signature of jamming in
  granular media},\ }\href@noop {} {\bibfield  {journal} {\bibinfo  {journal}
  {Nature}\ }\textbf {\bibinfo {volume} {435}},\ \bibinfo {pages} {1075}
  (\bibinfo {year} {2005})}\BibitemShut {NoStop}%
\bibitem [{\citenamefont {Mari}\ \emph {et~al.}(2009)\citenamefont {Mari},
  \citenamefont {Krzakala},\ and\ \citenamefont {Kurchan}}]{mari2009jamming}%
  \BibitemOpen
  \bibfield  {author} {\bibinfo {author} {\bibfnamefont {R.}~\bibnamefont
  {Mari}}, \bibinfo {author} {\bibfnamefont {F.}~\bibnamefont {Krzakala}},\
  and\ \bibinfo {author} {\bibfnamefont {J.}~\bibnamefont {Kurchan}},\
  }\bibfield  {title} {\bibinfo {title} {Jamming versus glass transitions},\
  }\href@noop {} {\bibfield  {journal} {\bibinfo  {journal} {Phys. Rev. Lett.}\
  }\textbf {\bibinfo {volume} {103}},\ \bibinfo {pages} {025701} (\bibinfo
  {year} {2009})}\BibitemShut {NoStop}%
\bibitem [{\citenamefont {Acharya}\ \emph {et~al.}(2020)\citenamefont
  {Acharya}, \citenamefont {Sengupta}, \citenamefont {Chakraborty},\ and\
  \citenamefont {Ramola}}]{acharya2020athermal}%
  \BibitemOpen
  \bibfield  {author} {\bibinfo {author} {\bibfnamefont {P.}~\bibnamefont
  {Acharya}}, \bibinfo {author} {\bibfnamefont {S.}~\bibnamefont {Sengupta}},
  \bibinfo {author} {\bibfnamefont {B.}~\bibnamefont {Chakraborty}},\ and\
  \bibinfo {author} {\bibfnamefont {K.}~\bibnamefont {Ramola}},\ }\bibfield
  {title} {\bibinfo {title} {Athermal fluctuations in disordered crystals},\
  }\href@noop {} {\bibfield  {journal} {\bibinfo  {journal} {Phys. Rev. Lett.}\
  }\textbf {\bibinfo {volume} {124}},\ \bibinfo {pages} {168004} (\bibinfo
  {year} {2020})}\BibitemShut {NoStop}%
\bibitem [{\citenamefont {Lerner}\ and\ \citenamefont
  {Bouchbinder}(2022)}]{lerner2022disordered}%
  \BibitemOpen
  \bibfield  {author} {\bibinfo {author} {\bibfnamefont {E.}~\bibnamefont
  {Lerner}}\ and\ \bibinfo {author} {\bibfnamefont {E.}~\bibnamefont
  {Bouchbinder}},\ }\bibfield  {title} {\bibinfo {title} {Disordered crystals
  reveal soft quasilocalized glassy excitations},\ }\href@noop {} {\bibfield
  {journal} {\bibinfo  {journal} {Phys. Rev. Lett.}\ }\textbf {\bibinfo
  {volume} {129}},\ \bibinfo {pages} {095501} (\bibinfo {year}
  {2022})}\BibitemShut {NoStop}%
\bibitem [{\citenamefont {Maharana}\ \emph {et~al.}(2024)\citenamefont
  {Maharana}, \citenamefont {Das}, \citenamefont {Chaudhuri},\ and\
  \citenamefont {Ramola}}]{maharana2024universal}%
  \BibitemOpen
  \bibfield  {author} {\bibinfo {author} {\bibfnamefont {R.}~\bibnamefont
  {Maharana}}, \bibinfo {author} {\bibfnamefont {D.}~\bibnamefont {Das}},
  \bibinfo {author} {\bibfnamefont {P.}~\bibnamefont {Chaudhuri}},\ and\
  \bibinfo {author} {\bibfnamefont {K.}~\bibnamefont {Ramola}},\ }\bibfield
  {title} {\bibinfo {title} {Universal stress correlations in crystalline and
  amorphous packings},\ }\href@noop {} {\bibfield  {journal} {\bibinfo
  {journal} {Phys. Rev. E}\ }\textbf {\bibinfo {volume} {109}},\ \bibinfo
  {pages} {044903} (\bibinfo {year} {2024})}\BibitemShut {NoStop}%
\bibitem [{\citenamefont {Ellenbroek}\ \emph {et~al.}(2006)\citenamefont
  {Ellenbroek}, \citenamefont {Somfai}, \citenamefont {van Hecke},\ and\
  \citenamefont {van Saarloos}}]{ellenbroek2006critical}%
  \BibitemOpen
  \bibfield  {author} {\bibinfo {author} {\bibfnamefont {W.~G.}\ \bibnamefont
  {Ellenbroek}}, \bibinfo {author} {\bibfnamefont {E.}~\bibnamefont {Somfai}},
  \bibinfo {author} {\bibfnamefont {M.}~\bibnamefont {van Hecke}},\ and\
  \bibinfo {author} {\bibfnamefont {W.}~\bibnamefont {van Saarloos}},\
  }\bibfield  {title} {\bibinfo {title} {Critical scaling in linear response of
  frictionless granular packings near jamming},\ }\href@noop {} {\bibfield
  {journal} {\bibinfo  {journal} {Phys. Rev. Lett.}\ }\textbf {\bibinfo
  {volume} {97}},\ \bibinfo {pages} {258001} (\bibinfo {year}
  {2006})}\BibitemShut {NoStop}%
\bibitem [{\citenamefont {Tong}\ and\ \citenamefont
  {Xu}(2014)}]{tong2014order}%
  \BibitemOpen
  \bibfield  {author} {\bibinfo {author} {\bibfnamefont {H.}~\bibnamefont
  {Tong}}\ and\ \bibinfo {author} {\bibfnamefont {N.}~\bibnamefont {Xu}},\
  }\bibfield  {title} {\bibinfo {title} {Order parameter for structural
  heterogeneity in disordered solids},\ }\href@noop {} {\bibfield  {journal}
  {\bibinfo  {journal} {Phys. Rev. E}\ }\textbf {\bibinfo {volume} {90}},\
  \bibinfo {pages} {010401(R)} (\bibinfo {year} {2014})}\BibitemShut {NoStop}%
\bibitem [{\citenamefont {Hexner}\ \emph {et~al.}(2019)\citenamefont {Hexner},
  \citenamefont {Urbani},\ and\ \citenamefont {Zamponi}}]{hexner2019can}%
  \BibitemOpen
  \bibfield  {author} {\bibinfo {author} {\bibfnamefont {D.}~\bibnamefont
  {Hexner}}, \bibinfo {author} {\bibfnamefont {P.}~\bibnamefont {Urbani}},\
  and\ \bibinfo {author} {\bibfnamefont {F.}~\bibnamefont {Zamponi}},\
  }\bibfield  {title} {\bibinfo {title} {Can a large packing be assembled from
  smaller ones?},\ }\href@noop {} {\bibfield  {journal} {\bibinfo  {journal}
  {Phys. Rev. Lett.}\ }\textbf {\bibinfo {volume} {123}},\ \bibinfo {pages}
  {068003} (\bibinfo {year} {2019})}\BibitemShut {NoStop}%
\bibitem [{\citenamefont {Shang}\ \emph {et~al.}(2025)\citenamefont {Shang},
  \citenamefont {Wang}, \citenamefont {Pan}, \citenamefont {Jin},\ and\
  \citenamefont {Zhang}}]{shang2025jamming}%
  \BibitemOpen
  \bibfield  {author} {\bibinfo {author} {\bibfnamefont {J.}~\bibnamefont
  {Shang}}, \bibinfo {author} {\bibfnamefont {Y.}~\bibnamefont {Wang}},
  \bibinfo {author} {\bibfnamefont {D.}~\bibnamefont {Pan}}, \bibinfo {author}
  {\bibfnamefont {Y.}~\bibnamefont {Jin}},\ and\ \bibinfo {author}
  {\bibfnamefont {J.}~\bibnamefont {Zhang}},\ }\bibfield  {title} {\bibinfo
  {title} {Jamming as a topological satisfiability transition with contact
  number hyperuniformity and criticality},\ }\href@noop {} {\bibfield
  {journal} {\bibinfo  {journal} {arXiv:2506.16474}\ } (\bibinfo {year}
  {2025})}\BibitemShut {NoStop}%
\bibitem [{\citenamefont {Torquato}(2018)}]{torquato2018hyperuniform}%
  \BibitemOpen
  \bibfield  {author} {\bibinfo {author} {\bibfnamefont {S.}~\bibnamefont
  {Torquato}},\ }\bibfield  {title} {\bibinfo {title} {Hyperuniform states of
  matter},\ }\href@noop {} {\bibfield  {journal} {\bibinfo  {journal} {Phys.
  Rep.}\ }\textbf {\bibinfo {volume} {745}},\ \bibinfo {pages} {1} (\bibinfo
  {year} {2018})}\BibitemShut {NoStop}%
\bibitem [{\citenamefont {Zachary}\ \emph {et~al.}(2011)\citenamefont
  {Zachary}, \citenamefont {Jiao},\ and\ \citenamefont
  {Torquato}}]{zachary2011hyperuniform}%
  \BibitemOpen
  \bibfield  {author} {\bibinfo {author} {\bibfnamefont {C.~E.}\ \bibnamefont
  {Zachary}}, \bibinfo {author} {\bibfnamefont {Y.}~\bibnamefont {Jiao}},\ and\
  \bibinfo {author} {\bibfnamefont {S.}~\bibnamefont {Torquato}},\ }\bibfield
  {title} {\bibinfo {title} {Hyperuniform long-range correlations are a
  signature of disordered jammed hard-particle packings},\ }\href@noop {}
  {\bibfield  {journal} {\bibinfo  {journal} {Phys. Rev. Lett.}\ }\textbf
  {\bibinfo {volume} {106}},\ \bibinfo {pages} {178001} (\bibinfo {year}
  {2011})}\BibitemShut {NoStop}%
\bibitem [{\citenamefont {Ikeda}\ \emph {et~al.}(2017)\citenamefont {Ikeda},
  \citenamefont {Berthier},\ and\ \citenamefont {Parisi}}]{ikeda2017large}%
  \BibitemOpen
  \bibfield  {author} {\bibinfo {author} {\bibfnamefont {A.}~\bibnamefont
  {Ikeda}}, \bibinfo {author} {\bibfnamefont {L.}~\bibnamefont {Berthier}},\
  and\ \bibinfo {author} {\bibfnamefont {G.}~\bibnamefont {Parisi}},\
  }\bibfield  {title} {\bibinfo {title} {Large-scale structure of randomly
  jammed spheres},\ }\href@noop {} {\bibfield  {journal} {\bibinfo  {journal}
  {Phys. Rev. E}\ }\textbf {\bibinfo {volume} {95}},\ \bibinfo {pages} {052125}
  (\bibinfo {year} {2017})}\BibitemShut {NoStop}%
\bibitem [{\citenamefont {Wu}\ \emph {et~al.}(2015)\citenamefont {Wu},
  \citenamefont {Olsson},\ and\ \citenamefont {Teitel}}]{wu2015search}%
  \BibitemOpen
  \bibfield  {author} {\bibinfo {author} {\bibfnamefont {Y.}~\bibnamefont
  {Wu}}, \bibinfo {author} {\bibfnamefont {P.}~\bibnamefont {Olsson}},\ and\
  \bibinfo {author} {\bibfnamefont {S.}~\bibnamefont {Teitel}},\ }\bibfield
  {title} {\bibinfo {title} {Search for hyperuniformity in mechanically stable
  packings of frictionless disks above jamming},\ }\href@noop {} {\bibfield
  {journal} {\bibinfo  {journal} {Phys. Rev. E}\ }\textbf {\bibinfo {volume}
  {92}},\ \bibinfo {pages} {052206} (\bibinfo {year} {2015})}\BibitemShut
  {NoStop}%
\end{thebibliography}
%

\section{End Matter}
\subsection{Simulation methods}
We study zero-temperature jammed packings of frictionless particles in two (2D) and three dimensions (3D) with periodic boundary conditions. Particles interact via finite-range repulsion
\begin{equation}
V(r_{ij})=\epsilon(1-r_{ij}/\sigma_{ij})^\alpha/\alpha
\end{equation}
for $r_{ij}\leq\sigma_{ij}$ and zero otherwise, where $r_{ij}$ is the distance between particles $i$ and $j$, and $\sigma_{ij}$ is the sum of their radii. $\alpha=2$ and $2.5$ correspond to harmonic and Hertzian systems.
The length unit is denoted as $\sigma$ and set by the average particle diameter for polydisperse systems and the small particle diameter for quenched binary systems. The mass, energy and pressure are in units of $m$, $\epsilon$, and $\epsilon/\sigma^d$, respectively.

To continuously modulate the structural order, packings were evolved from perfect crystalline structures (hexagonal in 2D with $\phi=0.92$, and face-centered cubic in 3D with $\phi=0.75$) by quasistatically increasing the particle-size polydispersity $\eta$ while keeping the packing fraction $\phi$ fixed \cite{tong2015crystals}.
To study how solids with distinct structural order behave when approaching the jamming transition, we modulate the packing fraction at fixed polydispersity to achieve pressures ranging from $p=2\times 10^{-2}$ to $5\times 10^{-7}$. To ensure perfect crystalline order, the simulation box aspect ratios are set to $L_y=\sqrt{3}L_x/2$ in 2D and $L_x=L_y=L_z$ in 3D.

We consider two types of particle-size distribution to examine their influence on the mechanism of self-organization. The diameter of particle $i$ is set as $\sigma_{i} = (1 + x_{i} \eta)\sigma$, with $\eta$ being the polydispersity. For the uniform distribution, $x_i$ is uniformly distributed within $[-0.5, 0.5]$; whereas for the Gaussian distribution, $x_i$ is drawn from a standard Gaussian distribution. Note that the random numbers $\{x_i\}$ are fixed during the change of polydispersity and pressure, and different sets of $\{x_i\}$ give independent realizations of jammed packings. In addition, the classical model of amorphous solids, i.e., randomly jammed packings of $50$:$50$ binary mixtures in an equilateral simulation box, is studied as a reference \cite{o2003jamming}. To be simple, all particles have the same mass.
Rattlers, i.e., particles that do not contribute to the rigidity of the system, are excluded from all analyses.
To ensure good statistics, results are averaged over many independent realizations: $5000$ for $N = 1024$ and $100$ for $N = 102400$.

The fast inertial relaxation engine algorithm (FIRE) is typically very efficient for the relaxation of particle systems~\cite{bitzek2006structural}. However, for near-crystalline packings with very small $\eta$, we find that the computational time to reach mechanical equilibrium tends to diverge near the jamming transition. Mechanically unstable states are reached even when the residual unbalanced force reaches $f_{\rm un}<10^{-30}$ for all particles (using quadruple precision), such that the Hessian matrix contains negative eigenvalues. This instability stems from particles with $0 <z \le d$ coordination neighbours, forming nearly collinear or coplanar geometries (typically, $z=2$ in 2D and $z=2$ or $3$ in 3D), due to the underlying crystalline structure. Such unstable particles can preclude the precise determination of the critical scalings and the coordination-number distribution near the jamming transition.

To ensure stable mechanical equilibrium, we develop a hybrid FIRE algorithm with particle self-activation and swaps. First, the system is relaxed using standard FIRE until $f_{\rm un}<10^{-12}$, i.e., force balance is realized to a good approximation. Then, we check for unstable particles with $0<z\le d$. If yes, we perform two local operations for all unstable particles: (i) a slight displacement toward the nearest non-contacting neighbour, and (ii) a diameter swap with the neighbour having the most different size. The above two procedures are repeated until all unstable particles are eliminated, ensuring that the system reaches a stable mechanical equilibrium.

\subsection{Characterizations of contact and density hyperuniformity}
We consider two measures of contact hyperuniformity \cite{hexner2018two}. The real-space fluctuation of the average coordination number within subsystems of size $L^d$ is measured by
\begin{equation}
\sigma^{2}_{Z}(L) = L^d \left\langle \left( Z_{L} - Z \right)^2 \right\rangle,
\end{equation}
where $Z_{L} = \sum_{j=1}^{N_L} z_j/N_{L}$ and $Z = \sum_{j=1}^{N} z_j /N$ are the average coordination number within the subsystem containing $N_L$ particles and the whole system, respectively. The angular brackets denote the ensemble average over subsystems in a given configuration as well as many independent realizations. The corresponding Fourier-space characterization is the structure factor of the coordination-number fluctuation
\begin{equation}
S_{Z}(q) = \frac{1}{N}\left\langle \left |  \sum_{j=1}^{N} \delta z_{j}e^{-i\vec{q} \cdot \vec{r}_{j}}  \right | ^2 \right\rangle,
\label{eq6}
\end{equation}
where $q$ is the amplitude of wave vector $\vec{q}$, $\delta z_{j}=z_j - Z$, and the average is over different realizations.

To further resolve the contribution from particles with different coordination numbers, we characterize the fluctuation of each component. The real-space fluctuation of the coordination-number distribution in subsystems of size $L^d$ is defined as
\begin{equation}
\sigma^{2}_{f_{z}}(L) = L^d \left\langle \left( f_{z}^{L} - f_{z} \right)^2 \right\rangle,
\end{equation}
where $f_{z}^{L}$ and $f_{z}$ are the fraction of particles with $z$ coordination neighbours within the subsystem and the whole system, respectively. The component-resolved structure factor of the coordination-number fluctuation is then defined as
\begin{equation}
S_{z=m}(q) = \frac{1}{N_{m}}\left\langle \left |  \sum_{j=1}^{N_{m}} \delta z_{j}e^{-i\vec{q} \cdot \vec{r}_{j}}  \right | ^2 \right\rangle,
\end{equation}
with the summation going over the $N_m$ particles with $z=m$.

We quantify the density fluctuation by the spectral density $\chi_{V}(q)$, which correctly takes into account the particle size dispersity \cite{torquato2018hyperuniform}
\begin{equation}
    \chi_{V}(q) = \frac{1}{V}\left\langle \phi(\vec{q}) \phi(-\vec{q}) \right\rangle,
\end{equation}
where $V$ is the system volume (or area in 2D) and the density field in Fourier space is given by
\begin{equation}
    \phi(\vec{q})=\sum_{j=1}^{N} J_{d/2}(qR_{j}) \left(\frac{2\pi R_{j}}{q}\right)^{d/2} e^{-i\vec{q} \cdot \vec{r}_{j}}.
\end{equation}
Here $R_{j}$ is the radius of particle $j$, $d$ is the spatial dimension, and $J_{n}(x)$ is the Bessel function of order $n$.

\end{document}